\documentclass[times,twocolumn,final,authoryear]{elsarticle}
\usepackage{jasr}
\usepackage{framed,multirow}

\usepackage{amssymb}
\usepackage{amsmath}
\usepackage{amsthm}
\usepackage{eurosym}
\usepackage{nicefrac}
\usepackage{imakeidx}

\usepackage{enumitem}

\usepackage[switch]{lineno}

\usepackage{float}
\usepackage{url}
\usepackage{xcolor}
\usepackage{soul}

\definecolor{newcolor}{rgb}{.8,.349,.1}

\usepackage[citebordercolor=white]{hyperref}

\usepackage{color}
\usepackage{tabularray}

\usepackage[normalem]{ulem}

\usepackage[
  input-uncertainty-signs=\pm,
  separate-uncertainty = true,
  group-separator = {\,},
	detect-weight=true,
  detect-family,
  print-unity-mantissa = false,
  parse-numbers=false
]{siunitx}
\DeclareSIUnit[]\km{\kilo\meter}
\DeclareSIUnit[]\mumeter{\micro\meter}
\DeclareSIUnit[]\um{\micro\meter}
\DeclareSIUnit[]\uHz{\micro\hertz}
\DeclareSIUnit[]\mHz{\milli\hertz}

\newcommand{\mpsssqrthz}[1]{\qty{#1}{\m\per\s\squared\per\sqrt{\text{Hz}}}}
\newcommand{\mpsssqrthzfrac}[1]{\frac{\qty{#1}{\m\per\s\squared}}{\qty{}{\sqrt{\text{Hz}}}}}
\DeclareSIUnit[per-mode=symbol]\nms{\nm\per\s\squared}

\usepackage{todonotes} 
\newcommand{\sasdf}{S^{\nicefrac{1}{2}} (f)}
\newcommand{\twopif}{2\pi f}

\renewcommand{\vec}[1]{\boldsymbol{#1}} 

\journal{Advances in Space Research}

\begin{document}

\verso{Alexey Kupriyanov \textit{et al.}}

\begin{frontmatter}

\title{Benefit of enhanced electrostatic and optical accelerometry for future gravimetry missions.\tnoteref{tnote1}}%

\tnotetext[tnote1]{}

\author[1]{Alexey \snm{Kupriyanov}\corref{cor1}}
\cortext[cor1]{Corresponding author: 
  Tel.: +49-511-762-5697}
\ead{kupriyanov@ife.uni-hannover.de}
\author[2,3]{Arthur \snm{Reis}}
\ead{arthur.reis@aei.mpg.de}
\author[4]{Manuel \snm{Schilling}}
\ead{Manuel.Schilling@dlr.de}
\author[2,3]{Vitali \snm{Müller}}
\ead{vitali.mueller@aei.mpg.de}
\author[1]{Jürgen \snm{Müller}}
\ead{mueller@ife.uni-hannover.de}

\address[1]{Institute of Geodesy, Leibniz University Hannover, Schneiderberg 50, Hannover 30167, Germany}
\address[2]{Max Planck Institute for Gravitational Physics (IGP), Albert Einstein Institute, Callinstraße 38, Hannover 30167, Germany}
\address[3]{Institute for Gravitational Physics, Leibniz University Hannover, Callinstraße 38, Hannover 30167, Germany}
\address[4]{Institute for Satellite Geodesy and Inertial Sensing, German Aerospace Center (DLR), Callinstraße 30B, Hannover 30167, Germany}

\received{19 May 2023}
\finalform{21 December 2023}
\accepted{31 December 2023}
\availableonline{3 January 2024}

\begin{abstract}

Twenty years of gravity observations from various satellite missions have provided unique data about mass redistribution processes in the Earth system, such as melting of Greenland’s ice shields, sea level changes, ground and underground water depletion, droughts, floods, etc. The ongoing climate change underlines the urgent need to continue this kind of observations with future gravimetry missions using enhanced concepts and sensors. This paper studies the benefit of enhanced electrostatic and novel optical accelerometers and gradiometers for future gravimetry missions.

One of the limiting factors in the current space gravimetry missions is the drift of the Electrostatic Accelerometers (EA) which dominates the error contribution at low frequencies (\qty{<1}{\mHz}). This study focuses on the modeling of enhanced EAs with laser-interferometric readout, so-called optical accelerometers, and on evaluating their performance for gravity field recovery in future satellite missions.

In this paper, we simulate gravimetry missions in multiple scopes, applying various software modules for satellite dynamics integration, accelerometer (ACC) and gradiometer simulation and gravity field recovery. The total noise budgets of the modeled enhanced electrostatic and optical ACCs show a similar sensitivity as the ACC concepts from other research groups. Parametrization w.r.t. the weight of the test mass (TM) of ACCs and the gap between the TM and the surrounding electrode housing confirmed the fact known from previous results that an ACC with a heavier TM and a larger gap will perform better. Our results suggest that the anticipated gain of novel ACCs might at some point be potentially limited by noise from the inter-satellite laser ranging interferometry.

In order to present the advantage of the novel sensors, time-variable background models and associated aliasing errors were not considered in our simulations. The utilization of enhanced EAs and optical ACCs shows a significant improvement of accuracy compared to the currently used GRACE-like EA. In addition, their benefit in double satellite pairs in a so-called Bender constellation as well as in the combination of low-low satellite-to-satellite tracking with cross-track gradiometry has been investigated.

%%%%
\end{abstract}

\end{frontmatter}

%\linenumbers

\section{Introduction}
\label{intro}

\subsection{Relevance of the research}
In the era of dedicated satellite gravimetry missions, starting in the beginning of the 21st century, the Earth system has experienced changes, either due to natural causes or because of direct human impact \citep{Flechtner2010,Rodell.2018}. Global warming and its diverse consequences, such as sea level rise, severe climate trends, floods and droughts, attracted the public's attention during the last decade. Scientific communities, decision makers and governments have shown a deep interest in satellite gravimetry missions to monitor the time-variable gravity field of the Earth, which provide unique information about the mass redistribution processes \citep{Pail2015a}. 
These processes are mostly driven by the hydrological cycle, such as seasonal water storage changes in the Amazon river basin \citep{Tourian.2018}, global warming effects, like ice sheet melting in Greenland and Antarctica ice sheet melting \citep{Siemes.2013, Velicogna.2020, Otosaka.2023}, and human activity, e.g., groundwater depletion in the Indian subcontinent \citep{Frappart.2018,Asoka.2020}.
Continuous measurements from the dedicated satellite gravimetry missions -- Challenging Minisatellite Payload (CHAMP) \citep{torge2023geodesy}, Gravity Field and Steady-State Ocean Circulation Explorer (GOCE) \citep{Bruinsma.2014, Flechtner.2021}, Gravity Recovery And Climate Experiment (GRACE) \citep{Tapley.2019,Chen.2022,Panet.2022} and Gravity Recovery And Climate Experiment - Follow On (GRACE-FO) \citep{Chen.2020,Ciraci.2020, Peidou.2022} -- measured the Earth's mass change phenomena at different temporal and spatial scales.

As a continuation of the current GRACE-FO mission, another satellite pair will be built in a German-US partnership named Mass Change Mission. It is planned to be launched in 2027-2028 to a polar orbit. This mission will employ a lone Laser Ranging Interferometer (LRI) \citep[][]{Haagmans.2020_MAGIC} as the inter-satellite ranging instrument, motivated by the results of GRACE-FO \citep{Haagmans2020}.

\begin{figure}
  \centering
  \includegraphics[width=0.5\textwidth]{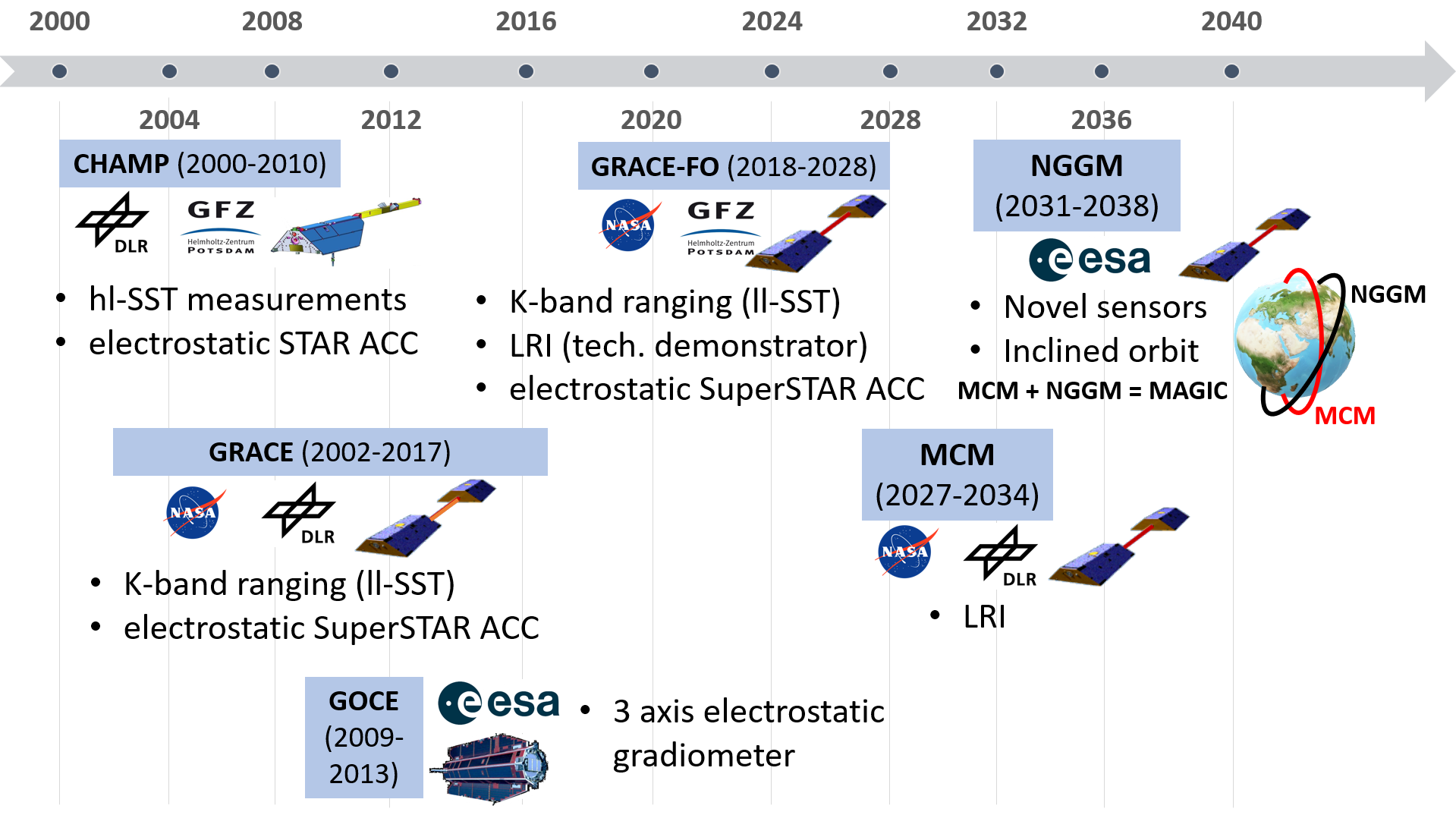}
  \caption{Timeline of the past, current and future gravimetry missions (modified from \citet[][]{Massotti2022})}
  \label{NGGM_timeline}
\end{figure}

The National Aeronautics and Space Administration (NASA) highlighted mass change missions as one of its top five priorities \citep{Wiese2022}. In November 2022, the European Space Agency (ESA) Council Meeting at Ministerial Level allocated \EUR{2.7} billion for Earth science research, in particular \EUR{120} million for the \emph{Mass change And Geosciences International Constellation} (MAGIC) project \citep{ESA_2022}. This project should consist of two pairs of satellites: one pair in a polar orbit -- the Mass Change Mission by German Aerospace Center (DLR) and NASA mentioned above -- and one pair in an inclined orbit -- the Next Generation Gravimetry Mission (NGGM) by ESA \citep[][]{Massotti.2021}. Fig.~\ref{NGGM_timeline} shows the timeline of past and planned gravimetry missions with their major technical features. Additionally, India and China are also interested in satellite gravimetry and already develop their own gravimetry missions. China launched its technology demonstrator gravitational wave detector satellite TianQin-1 in 2019 and plans the launch of TianQin-2 to investigate technologies to be used in future gravimetry missions \citep{Luo2020}. Moreover, at the end of 2021, China successfully launched its first pair of gravimetry satellites to a low Earth orbit that utilizes high-low and low-low satellite-to-satellite tracking techniques. \citet[][]{Xiao2023} demonstrated that the design requirements for the satellite platform were met and that it is capable of characterizing global hydrological changes that are in agreement with the outputs of the GRACE-FO mission.

\subsection{Low-low satellite-to-satellite gravimetry principle}
\label{subsec:space_gravimetry}

\begin{figure}
    \centering
    \includegraphics[width=0.5\textwidth]{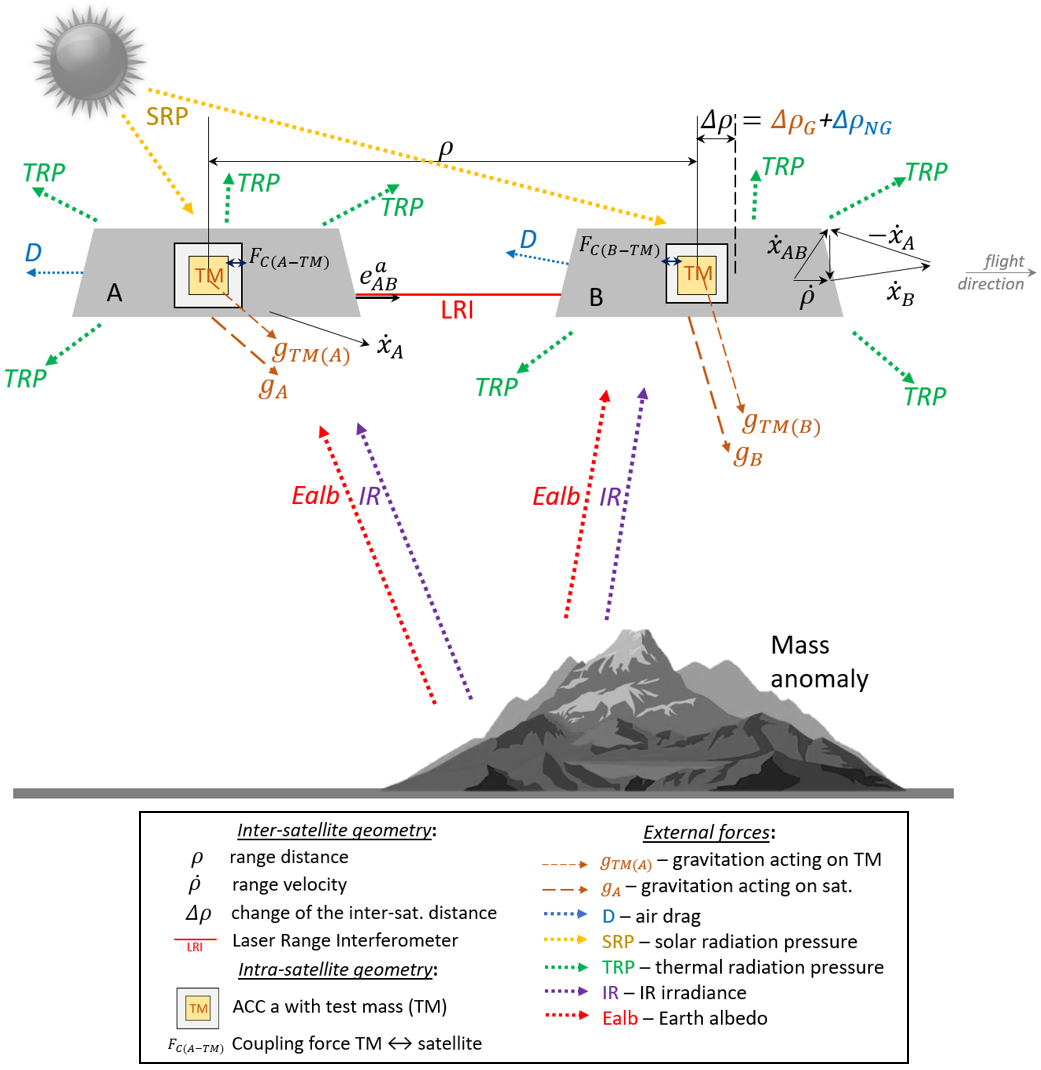}
    \caption{Geometry and measurement principle of ll-SST tracking of the GRACE-FO mission}
    \label{fig:grace_principle}
\end{figure}

Two representative satellite gravimetry missions are GRACE and GRACE-FO. Each used a pair of in-line satellites separated by about \qty{220}{\km} in an orbit with an inclination of \qty{89}{\degree} and an initial altitude of about \qty{450-500}{\km}, which decayed during the mission's lifetime. Both missions utilize the low-low satellite-to-satellite tracking (ll-SST) principle, depicted in Fig.~\ref{fig:grace_principle}, to measure the variations in the inter-satellite distance \citep{Daras.2015}. The satellites are accelerated differently due to the variations in the gravitational field and the differences in the non-gravitational forces acting on each satellite. 
Due to the difference of the gravitational ($\Delta \rho_G$) and non-gravitational ($\Delta \rho_{NG}$) forces acting on the individual satellite, the inter-satellite distance changes by $\Delta \rho$. This was measured by K-band ranging (KBR) in the GRACE and GRACE-FO missions, while the latter also successfully applied an LRI as a technology demonstrator. 
To derive the gravitational contribution in the distance variations, the inter-satellite ranging signal must be corrected for the non-gravitational contributions consisting of air drag, solar radiation pressure, thermal radiation pressure, Earth's albedo and IR irradiance, which are measured by accelerometers (ACCs) located at the center of mass (CoM) of each satellite.

\subsection{Limitations in current gravimetry missions}

One of the limiting factors in the current space gravimetry missions, in particular dominating the error contribution at low frequencies, is the Electrostatic Accelerometer's (EA) drift \citep[][]{Christophe2015}. This type of ACC measures the position of the test mass (TM) capacitively and keeps it centered in the electrode housing (EH) by electrostatic actuation forces. Figure~\ref{ASD_general_comparison} depicts the noise in terms of Amplitude Spectral Densities (ASD) from different types of ACCs together with a typical non-gravitational signal and the time-variable gravity field signal in the inter-satellite range acceleration. The EA used in GRACE-FO (blue curve) has a limited sensitivity in frequencies below $ \qty{1}{\milli\hertz}$. Its ASD is described as \citep[][]{Daras2017}: 
\begin{figure}
  \centering
  \includegraphics[width=0.5\textwidth]{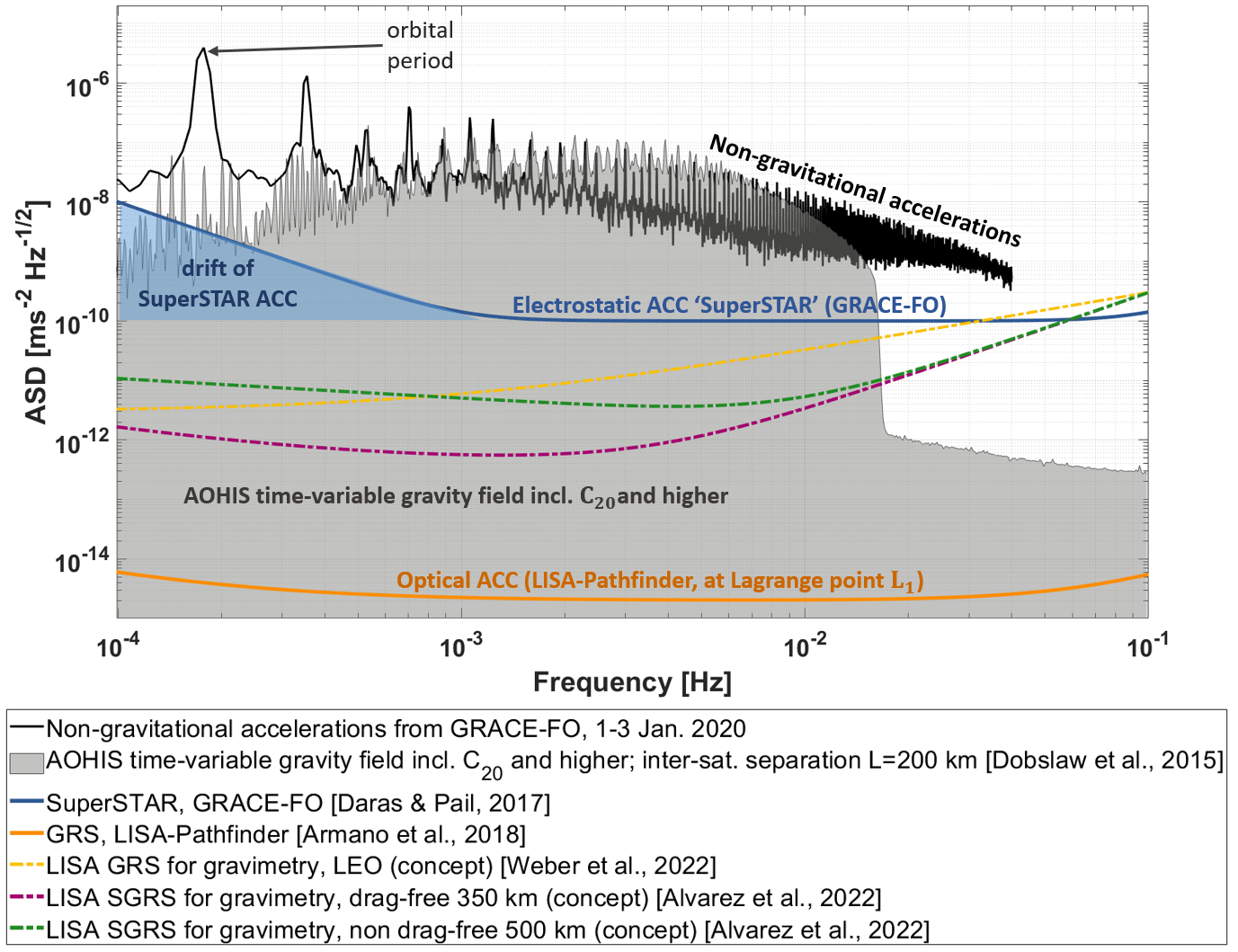}
  \caption{Comparison of the accelerometers ASD sensitivities for current instruments and advanced concepts w.r.t. typical non-gravitational accelerations (black trace). These non-gravitational accelerations need to be removed from the inter-satellite range observations by means of the ACC in order to retrieve the time-variable components of the gravity field (grey area)}
  \label{ASD_general_comparison}
\end{figure}
\begin{linenomath}\begin{multline}
  \sasdf = 10^{-10} \mpsssqrthzfrac{} 
  \sqrt{\frac{\left({\frac{\qty{1}{\mHz}}{f}}\right)^4}{\left({\frac{\qty{10}{\uHz}}{f}}\right)^4+1} + 1 + \left( \frac{f}{\qty{0.1}{\Hz}}\right)^4} \text{.}
   \label{eqn:ASD superSTAR EA [Daras2017]} 
\end{multline}\end{linenomath}

The LISA-Pathfinder (LPF) mission, which was operated in the quiet space environment of the Lagrange point $L_1$, had an optical ACC on board dubbed Gravitational Reference Sensor GRS; Fig.~\ref{ASD_general_comparison}, orange curve) \citep[][]{Armano.2018}. Contrary to electrostatic accelerometers, which measure the TM displacement capacitively and actuate it electrostatically, optical ones track the TM with laser interferometry. Its ASD is given by:

\begin{linenomath}\begin{multline}
    \sasdf = \mpsssqrthz{10^{-10}} ~ \\
    \times \left[ 2\times 10^{-5} + 3.5\times10^{-3}\left(\frac{f}{\qty{1}{\Hz}}\right)^2 + 5.8 \times 10^{-10} \left(\frac{\qty{1}{\Hz}}{{f}}\right)^{1.21} \right] \text{.}
    \label{eqn:ASD optic ACC, LPF [Armano et al., 2018]}
\end{multline}\end{linenomath}

Achieving such an accuracy was a real breakthrough and was even below the technical requirements of the future LISA mission. 
However, there is significant attenuation of the long-wavelength components of the gravity signal with altitude, forcing gravimetry missions to be in low Earth orbits, e.g., at altitudes \qty{<500}{\km} where non-gravitational forces are substantial.
In a low Earth orbit, it is not expected to reach a similar performance as demonstrated in LPF. 

\citet[][]{Alvarez2022} proposed the Simplified Gravitational Reference Sensor (SGRS): an advanced EA without optical TM position readout to be used in a drag-free mode in low Earth orbits at \qty{350}{\km} altitude (Fig.~\ref{ASD_general_comparison}, purple curve) and in non-drag free mode at an altitude of \qty{500}{\km} (green curve). The ASD of the mentioned ACC concept, which includes the contribution of the capacitive sensing dominating at the high frequency domain \citep[c.f. Fig.~13 of][]{Alvarez2022}, for a drag-free regime at \qty{350}{\km} altitude, is

\begin{linenomath}\begin{multline}
    \sasdf = 10^{-10}\mpsssqrthz{} ~\\
    \times \left[3\times 10^{2} \left(\frac{f}{\qty{1}{\Hz}}\right)^2 + 4 \times 10^{-3} \sqrt{1+\frac{\qty{700}{\uHz}}{f}+\left(\frac{\qty{300}{\uHz}}{f}\right)^2}\right] \text{,}
    \label{Alvarez_ASD_350}
\end{multline}\end{linenomath}

and for a non-drag-compensated \qty{500}{\km} altitude,  it is

\begin{linenomath}\begin{multline}
   \sasdf = 10^{-10}\mpsssqrthz{} ~ \\
   \times\left[3 \times 10^{2} \left(\frac{f}{\qty{1}{\Hz}}\right)^2 + 5 \times 10^{-3} \sqrt{1+\left(\frac{\qty{1}{\Hz}}{f}\right)^{2/3}}\right]\text{.}
    \label{Alvarez_ASD_500}
\end{multline}\end{linenomath}

Another LISA-derived ACC concept for future gravimetry missions was analyzed by \citet{Weber2022} \citep[c.f. Fig.~1 of][and yellow curve in Fig.~\ref{ASD_general_comparison}]{Weber2022} with an ASD of

%\begin{linenomath}\begin{equation}
    %\sasdf =  10^{-10} \cdot \left[ 3\times 10^{-2}+ \frac{30 f}{\qty{1}{\Hz}}  \right] \mpsssqrthzfrac{} \text{.}
%\end{equation}\end{linenomath}
\begin{linenomath}\begin{equation}
    \sasdf = \left[ 3\times 10^{-2}+ \frac{30 f}{\qty{1}{\Hz}}  \right]10^{-10} \mpsssqrthzfrac{} \text{.}
\end{equation}\end{linenomath}
\subsection{Research goals and paper outline}

This work aims to investigate new payload concepts for future satellite gravimetry missions through simulations, for either single- or double-pair constellations and with novel ACCs and gradiometers. One idea is to transfer as much knowledge as possible from the LISA-Pathfinder gravitational wave detector demonstrator mission into gravimetry missions at low Earth orbits, in particular in the fields of optical readout of the TM position, wire-free TM discharge and reduction of stiffness between TM and spacecraft. 

This study focuses on the evaluation of the potential benefit of novel instruments, such as the ones proposed by \citet[][]{Douch2017}, \citet[][]{Alvarez2022} and \citet[][]{Weber2022} to be used as ACCs or in gradiometers. Since the short-term mass variations in the atmosphere and ocean produce so-called aliasing errors, which are typically the main contributors to inaccuracies in GRACE solutions, the Atmosphere and Ocean De-Aliasing Level-1B products are constantly being improved. The latest releases are, e.g., release RL06 by \citet[][]{Dobslaw.2017} and release RL07 by \citet[][]{Shihora.2022}. However, necessary developments in the background models will not be addressed in this work. So further investigation of the full potential of new sensor technologies would be possible if the modeling for the reduction of temporal aliasing effects is improved. 

This paper is organized as follows: after an introduction Section~\ref{intro}), Section~\ref{sec:workflow} briefly explains the simulation procedure and the used software. In Section~\ref{Creating various accelerometers models} the procedure of ACC modeling within MATLAB/Simulink is described, followed by Section~\ref{sec: Gradiometers modelling}, where mathematical considerations regarding gradiometer modeling are presented. Gravity field recovery (GFR) simulations for a selection of mission scenarios, concepts and sensors are presented in Section~\ref{sec:GFR}. Finally, the conclusions and acknowledgments are shown in Sections~\ref{sec:conclusions} and \ref{sec:Acknowledgments}.

\section{Procedure for simulations and gravity field recovery}
\label{sec:workflow}

\begin{figure}
  \centering
  \includegraphics[width=0.5\textwidth]{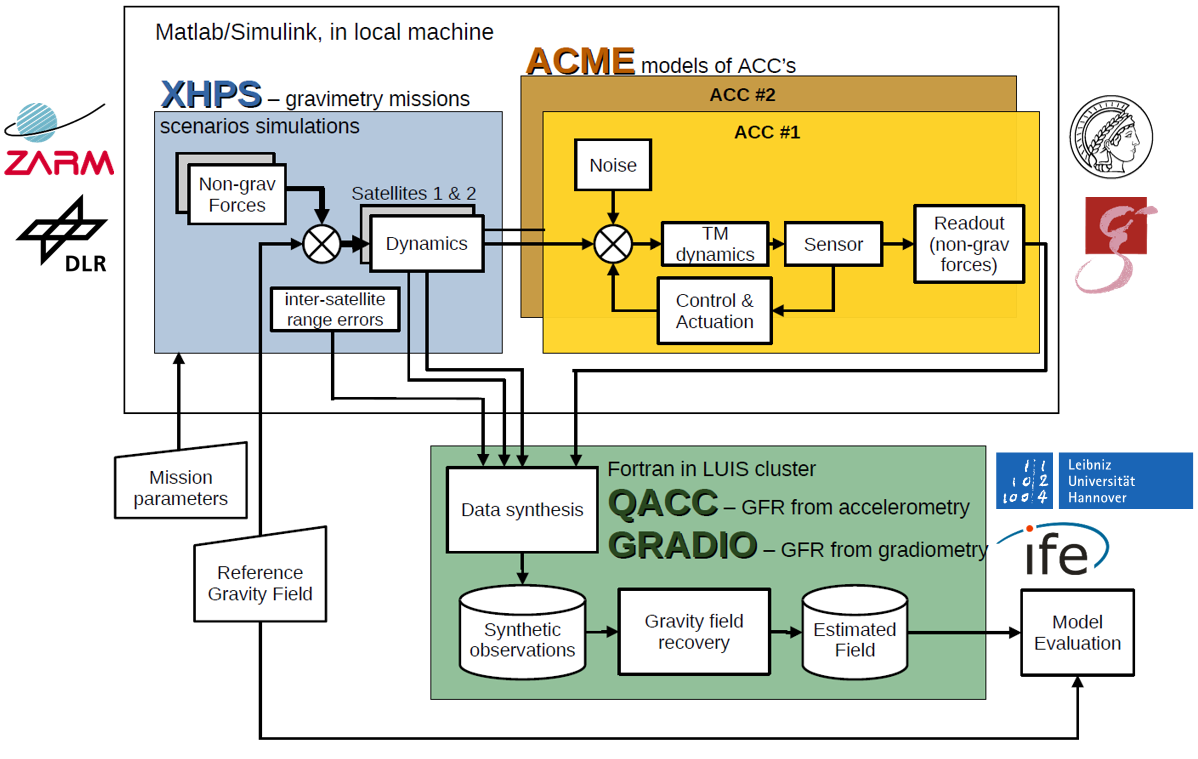}
  \caption{Block diagram of simulation procedure within the used software parts} 
  \label{B01_blockdiagram}
\end{figure}

This section introduces the workflow of the simulation procedure, including the used software and gives a short description of the used methods and approaches in the toolboxes. A block diagram of the simulation procedure of the various software parts is shown in Fig.~\ref{B01_blockdiagram}.

The dynamics of the satellite pair were simulated in the eXtended High Performance Satellite Dynamics Simulator (XHPS; blue box in Fig.~\ref{B01_blockdiagram})  in Matlab/Simulink \citep[][]{Woske_F}, considering several mission scenarios, e.g., a GRACE-like polar satellite pair or a double-satellite pair in a Bender constellation. This software was developed by the Center of Applied Space Technology and Microgravity (ZARM, Bremen) and is currently maintained by the Institute for Satellite Geodesy and Inertial Sensing, German Aerospace Center (DLR, Bremen). The flight dynamics are simulated taking into account a set of non-gravitational forces and a high-fidelity reference gravity field model. Non-gravitational forces that are used in XHPS were imported from corresponding databases and are calculated at each time step of the integrated orbit, based on a detailed finite element model of the satellite \citep[][]{Woske.2018}. An explanation of the double-pair satellite configuration is presented in Section~\ref{subsec GFR Bender}.

In order to model any type of both electrostatic and optical ACCs which utilizes optical interferometry to readout the TM position, a framework called Accelerometer Modeling Environment (ACME; yellowish boxes in Fig.~\ref{B01_blockdiagram}) was developed in Matlab/Simulink. This software was mainly developed by the Max Planck Institute for Gravitational Physics at Albert Einstein Institute, in the project B01 of the SFB 1464 TerraQ (Relativistic and Quantum-based Geodesy) \emph{`New measurement concepts with laser interferometers'}, and it was tested and integrated into XHPS in collaboration with the Institute of Geodesy at the Leibniz University Hannover. 

Different sensor types, noises and a wide range of other parameters can be included in the model of the ACC. ACME was designed to operate in two different modes: as a `standalone bench' for simulating the behavior of different ACCs, and as an integration module, which can be embedded in XHPS to simulate an ACC in an orbiting regime. Past, current and future gravimetry missions were simulated using `classical' (electrostatic ACC and KBR) or `novel' (enhanced electrostatic or optical ACC and LRI) sensors.

After simulating realistic observations with sensor models, GFR was performed by two software packages, programmed in Fortran: Quantum Accelerometry (QACC) and GRADIO, both developed at the Institute of Geodesy at Leibniz University Hannover \citep[][]{Wu2016} and run at the Leibniz University IT Service computational cluster due to their high computational demands. Their workflow is identical and explained in detail in subsection \ref{subsec:gfr_qacc_gradio}. In short, noise time-series representing ACCs (from ACME) and inter-satellite range sensor errors are added to the noise-free range rate and range acceleration time series (from XHPS). The synthesized data is used to estimate gravity field parameters of different future gravimetry mission concepts and mission constellations. Then, the obtained spherical harmonic (SH) coefficients are compared with the reference gravity field model. The QACC software recovers the gravity field for ll-SST cases (GRACE type) while GRADIO does it for gradiometry cases (GOCE type).

\subsection{Satellites dynamics simulator in XHPS}
\label{subsec:xhps}
XHPS is a MATLAB/Simulink toolbox that allows to simulate various satellite gravimetry mission scenarios. It calculates satellite dynamics via its own implementation of a multistep predictor-corrector integrator \citep[][]{Woske_F}, taking into account selected gravity field models of the Earth, tidal effects, third-body contributions, and also non-gravitational forces. It conveniently combines data in many frames, such as Earth Centered Inertial, Earth Centered Earth-Fixed and Satellite Body Fixed frame \citep[][]{Woske_F, woske2016development}.
The program generates a time series of the satellites' state-vectors, composed of positions, velocities, orientation quaternions and angular velocities of the satellite(s), either in Earth Centered Inertial or in the Earth-Centered Earth-Fixed reference frames, which are necessary for GFR.

\subsection{Accelerometer modeling within ACME}

ACME is a MATLAB/Simulink toolbox that simulates the behavior of the ACC and includes noise models of sensors (capacitive, optical), actuators (electrostatic) and other subsystems. 

Also, it provides the frequency response of the linearized system or runs time-domain simulations. The latter ones can be used concurrent to XHPS’s orbital dynamics simulation, effectively generating mock data of a flight-like instrument for gravity field recovery or evaluation of the spacecrafts' control schemes.

\subsection{Gravity field recovery with QACC and GRADIO}
\label{subsec:gfr_qacc_gradio}

The primary observable of GRACE-like missions is the inter-satellite range change due to the Earth gravity field and other forces -- measured with KBR and/or LRI. A number of approaches link the observed range accelerations to the gravitational potential. In general, these approaches can be split in two major groups: time-wise and space-wise. The GFR software packages used in this work, QACC and GRADIO, both written in Fortran, utilize the acceleration approach that belongs to the space-wise group and will be discussed in more detail.

As illustrated in Fig.~\ref{B01_blockdiagram}, noise-free data (position and velocity) from the XHPS simulation with the state-vector of each satellite for each time step is provided as input to the adequate GFR software. Noise-free orbital data means that no GPS noise was used in its generation. It is worth noting that the effect of the centrifugal term was neglected here in this study, since the focus is put on the benefit of novel sensors. From this data, noise-free range rate and range acceleration time series are generated in the `synthesis' module of the GFR software. Afterwards, noise time-series representing ACC and inter-satellite range instrument errors are added to the noise-free ranging time-series. The `recovery' module carries out a least squares adjustment and the resulting residuals are further used for the calculation of the variance-covariance matrix. At last, the `Evaluation' module validates the retrieved gravitational field model.  

In \citet{weigelt2017acceleration}, an equation of motion for a GRACE-like case was given as
\begin{linenomath}\begin{equation}
    \nabla \vec{V}_{AB} \cdot \vec{e}_{AB}^a = \ddot{\vec{\rho}} -  \frac{1}{\vec{\rho}} \left( \dot{\vec{x}}_{AB} \cdot \dot{\vec{x}}_{AB} - \dot{\vec{\rho}}^2\right)\text{,}
    \label{GRACE_eq_motion}
\end{equation}\end{linenomath}
where $\vec{\rho}$ is the observed (or simulated) range between the satellites A and B, $\vec{x}_{AB}$ - relative position, $\vec{e}_{AB}^a$ - unit vector along line of sight (LOS) between two spacecraft, $\dot{\vec{x}}_{AB}$ - relative velocity vector,  $\dot{\vec{\rho}}$ - range velocity and $\ddot{\vec{\rho}}$ - range acceleration (see also Fig.~\ref{fig:grace_principle}).
Equation~\eqref{GRACE_eq_motion} is the basic equation of the acceleration approach and connects the range acceleration ($\ddot{\vec{\rho}}$) to the gradient of the gravitational field ($\nabla V_{AB}$). 

The gravitational potential ${V}$ can be represented in a spherical harmonic series \citep[][]{hofmann2006physical} as
\begin{linenomath}\begin{multline}
    {V}(r,\theta, \lambda) = \frac{GM}{R}\sum_{n=0}^{\infty} \left( \frac{R}{r}\right)^{n+1} \\ \times \sum_{m=0}^{n} \left[ \overline{C}_{nm} \cos{m\lambda}+\overline{S}_{nm} \sin{m\lambda}\right]\overline{P}_{nm}(\cos{\theta})\text{,}
    \label{eqn_grav_potential}
\end{multline}\end{linenomath}

where $GM$ - gravitational constant of the Earth, $R$ - Earth's reference ellipsoid equatorial radius, $\left(r, \theta, \lambda \right)$ - spherical coordinates of a point on the Earth's surface, $n,m$ - SH degree and order, $\overline{P}_{nm}\left( \cos{\theta}\right)$ - fully normalized associated Legendre functions and $\overline{C}_{nm}, \overline{S}_{nm}$ - cosine and sine normalized SH coefficients, which are the unknowns of the gravity field solution.

The present-day satellite gravimetry missions provide a huge amount of observations, including thousands of parameters. For example, GFR up to degree and order 90 gives 4183 estimated coefficients (starting from $C_{20}$). This data forms a large-scale and over-determined linear equation system which can be solved by the least squares technique that was discussed in the context of the software packages by \citet{Wu2016} in detail. 

Accuracy and correlation of the measurements are represented by a stochastically modeled full variance-covariance matrix $\vec{\Sigma}_{ll}$:

\begin{linenomath}\begin{equation}
    \vec{\Sigma}_{ll} = \begin{bmatrix} 
    \sigma_{1}^2 & \sigma_{12} & \dots  & \sigma_{1n}\\
    \sigma_{21} & \sigma_{2}^2 & \dots  & \sigma_{2n}\\
    \vdots & \vdots & \ddots  & \vdots\\ 
    \sigma_{n1} & \sigma_{n2} & \dots  & \sigma_{n}^2
    \end{bmatrix}
\end{equation}\end{linenomath}

where $\sigma_{i}^2$ - variance of the $i$-th element and $\sigma_{ij}$ is the covariance between $i$-th and $j$-th measurement. The formulation to estimate the unknown solution $\hat{x}$ using least squares adjustment is given by \citet{koch1999parameter}. 

Post-fit residuals can be used to detect outliers in the observations. The Power Spectral Density of the residuals shows the spectral behavior of the measurement error. The empirical variance-covariance matrix is computed from the residuals and they are also used in the calculation of the posterior variance of the unit weight $\hat{\sigma}_{0}^2$, which gives the quality of the solution. The computation of the posterior variance is done by

\begin{linenomath}\begin{equation}
    \hat{\sigma}_{0}^2 = \frac{\vec{l}^T\vec{P}\vec{l}-\vec{W}^T\hat{\vec{x}}}{s-r} \text{,}
    \label{posterior_variance}
\end{equation}\end{linenomath}

where $l$ -- vector of observations, $P$ -- weight matrix, $W=A^TPl$, $A$ -- design matrix, $s$ -- number of observations and $r$ -- number of parameters. Such a least squares adjustment is used in our GFR software for ll-SST (QACC) and gradiometry (GRADIO).

\subsection{Validation of the retrieved gravity field models}
The retrieved gravity field can be validated in the spatial domain by computing the difference to a reference gravity field model and plotting the result on a global map, e.g., in terms of equivalent water height (EWH) \citep[][]{Wu2016}. It shows the geographical distribution of the errors. 

In the spectral domain, the validation of the retrieved gravity field can be based on the formal errors \citep[][]{Dahle.2019}. The sum of the squares of the SH coefficients at the same degree gives the total power of the coefficients. Often, the error degree variance is visually represented in terms of geoid heights.

\section{Creating various accelerometer models}
\label{Creating various accelerometers models}

\begin{figure}
  \centering
  \includegraphics[width=0.5\textwidth]{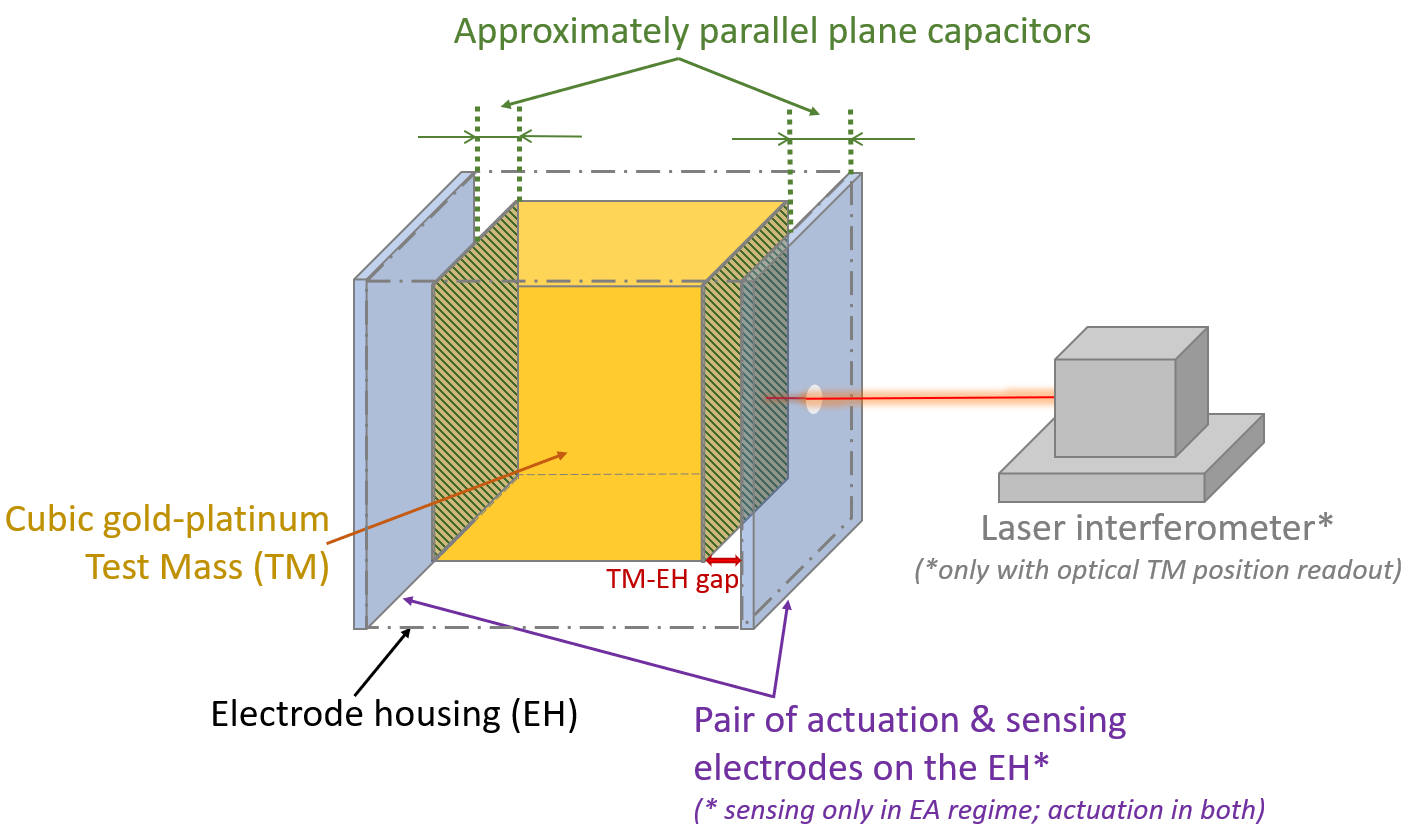}
  \caption{Illustration of 1 degree of freedom accelerometer model}
  \label{optical_ACC2}
\end{figure}

The core of the ACC model is the dynamics of the TM with mass $m$, (like the one represented in Fig.~\ref{optical_ACC2}) with respect to the satellite of mass $M$. The position and velocity vectors of the TM and satellite, as written in the inertial reference frame $I$, are $\vec{x}^I$, $\vec{\dot{x}}^I$, $\vec{X}^I$ and $\vec{\dot{X}}^I$. As previously mentioned, both the spacecraft and TM are affected by the gravitational acceleration $\vec{g}^I$. The spacecraft is also affected by a non-gravitational force  $\vec{F}_\text{NG}^I$, which includes atmospheric drag, solar and earth radiation pressure, etc. In the case of drag-free missions, the spacecraft is propelled by the force $\vec{F_\text{Thr}}^I$ of its thrusters. The TM and the spacecraft are coupled with a force $\vec{F_{C}}^I(\vec{X}^I,\vec{\dot{X}}^I,\vec{{x}^I},\vec{\dot{x}}^I)$. To keep the TM centered in its housing, it is actuated by a control force $\vec{F_\text{ctrl}}$ from electrodes, which are fixed with respect to the satellite.

Gravimetry missions such as GRACE only employ a single ACC, gradiometry missions apply multiple ones. Therefore, the full generic equations of movement should include $n$ TMs. However, for a ll-SST mission with a single TM located in the satellite's CoM ($\vec{x}^I\approx\vec{X}^I$), the equations become (adapted from \citet[][]{Theil_S.})
\begin{linenomath}\begin{multline}
    M{\vec{\ddot{X}}}^I = M\vec{g}^I(\vec{X}^I) + \vec{F_\text{NG}}^I(\vec{X}^I,\vec{\dot{X}}^I)  + \vec{F_\text{ctrl}}^I\\
    \qquad\qquad +  \vec{F_\text{Thr}}^I + \vec{F_{C}}^I(\vec{X}^I,\vec{\dot{X}}^I,\vec{x}^I,\vec{\dot{x}}^I) 
\end{multline}\end{linenomath}
   
and
    \begin{linenomath}\begin{equation}
        m{\vec{\ddot{x}}}^I = m\vec{g}^I(\vec{x}^I) + \vec{F_{C}}^I(\vec{x}^I,\vec{\dot{x}}^I,\vec{X}^I,\vec{\dot{X}}^I) + \vec{F_\text{ctrl}}^I\text{.}
    \end{equation}\end{linenomath}

In this type of mission, and without drag compensation, the equation of motion for a single TM in the satellite frame $S$ is simplified to 
\begin{linenomath}\begin{equation}
   m\ddot{\vec{x}}^{S} \approx \vec{F}_{C}^{S} - \vec{F_\text{ctrl}}^{S} \text{.} 
\end{equation}\end{linenomath}

The complete system of equations should account for all six degrees of freedom and their possible cross terms. In this study, however, the assumption is made that the relative displacements and inclinations are small and all non-linear, and the cross terms will be disregarded here.

The next step is to consider how the acceleration can be measured, both to feedback the control loop -- either for drag compensation or TM centering --  and to generate the data that is passed to the GFR software.
All techniques used so far in real instruments and modeled in our simulator use the measurement of the control signal voltage, which results in the actuation force needed to re-center the TM with respect to its housing and the appropriate transfer function to derive the acceleration. These techniques are either capacitive sensing or optical readout. The latter being subdivided into laser interferometry, optical levers and shadow sensors.

In the case of a capacitive sensor, the surface of the TM must be conductive and electrode plates are installed in the walls of the housing. With this arrangement, each electrode forms a capacitor with the nearest face of the TM. Together with its projected `shadow' on the surface of the TM, they can be considered as a plane parallel capacitor -- at least as a first approximation -- to derive the equations of the capacitive sensing and of the electrostatic actuation \citep{Josselin1999}. As the TM moves inside the housing, the distance between the electrode fixed in the housing and its shadow changes, and so does the capacitance value. A readout circuit measures this value to determine the control voltage, from which the acceleration can be inferred. 

For both the interferometry readout and the optical lever, the surface of the TM is a mirror and its displacement will cause a difference in the interference pattern or in the position of the beam in a quadrant photodiode, respectively \citep{huarcaya2020, Heinzel2003}. Lastly, shadow sensors, such as the one proposed to be used in  \citet{zoellner2013}, measure the power in a photodiode from a beam that is partially blocked by the moving TM, so that more or less light incides on the sensor.

Non-drag free missions require active control of the TM position, to keep it centered in its housing as the spacecraft orbits under the effect of non-conservative forces. While drag-free or drag-compensated missions can have in principle very low TM control authority in the measurement bandwidth during nominal science mode, control forces are still needed during commissioning and calibration phases.

The actuation is achieved by means of electrostatic repulsion. The TM is polarized with a voltage $U_p$, either by a connected wire or via capacitive injection. Multiple electrodes are installed in the interior faces of the housing, each with an area $A$, and fed with actuation voltage $U$ by a controller. The force from one electrode is a non-linear function of voltage and separation, however, if we consider the TM with no displacement ($x=0$) from its nominal position with a separation $d$ from the walls of the housing, the force can be written as
\begin{linenomath}\begin{equation}
    \vec{F}_\text{ctrl}(U) = \frac{2\epsilon A}{d^2}U_p U \vec{\hat{n}}
\end{equation}\end{linenomath}

where $\epsilon$ - the dielectric constant of vacuum and $\vec{\hat{n}}$ the vector normal to the electrode.

Our research is partially motivated by the very promising results of the LPF mission, which has demonstrated the benefit of using ultraviolet radiation charge management \citep[][]{Armano.2018b}. Electrical charge builds up on the TM over time, caused by cosmic ray bombardments and was dealt with in previous ACCs by connecting a thin soft wire, through which an arbitrary electric potential could be maintained. This, however, had the disadvantage of introducing a significant source of stiffness.
The wireless charge management system impinges ultraviolet light on the TM, which excites and expels the extra electrons at a rate that keeps the electrostatic sources of noises at an acceptable level \citep{sumner2020}. 

The sensitivity of the instrument model depends on the choice of measurement technique.
The capacitive sensing noise in turn depends on the electronics involved in the capacitance sensor. In circuits with differential transformer, e.g., one dominant noise source is the thermal noise of the sensing bridge \citep{Mance2012}. Circuits using differential instrumentation amplifiers are dominated by voltage noise from the charge amplifiers \citep{Lotters1999,Alvarez2022}. The sensing noise of a laser interferometric readout is below the level of non-gravitational accelerations, coupled through the stiffness for the frequency ranges of interest, and can be disregarded at first approximation \citep[][]{Alvarez2022}.

The thermal noise in an ACC is comprised of the Brownian motion of the residual gas in the enclosure and the differential pressure that arrives from unbalanced radiation, gas-wall collisions and outgassing in a near-isothermic environment \citep{carbone2007}. Actuation noise comes from the stability of the voltage reference used by the drivers for the actuation electrodes \citep[][]{Mance2012}. The non-gravitational forces that impact the satellite also affect the TM through the spacecraft -- TM coupling \citep[][]{Josselin1999}.

\begin{figure}
\centering
\includegraphics[width=.5\textwidth]{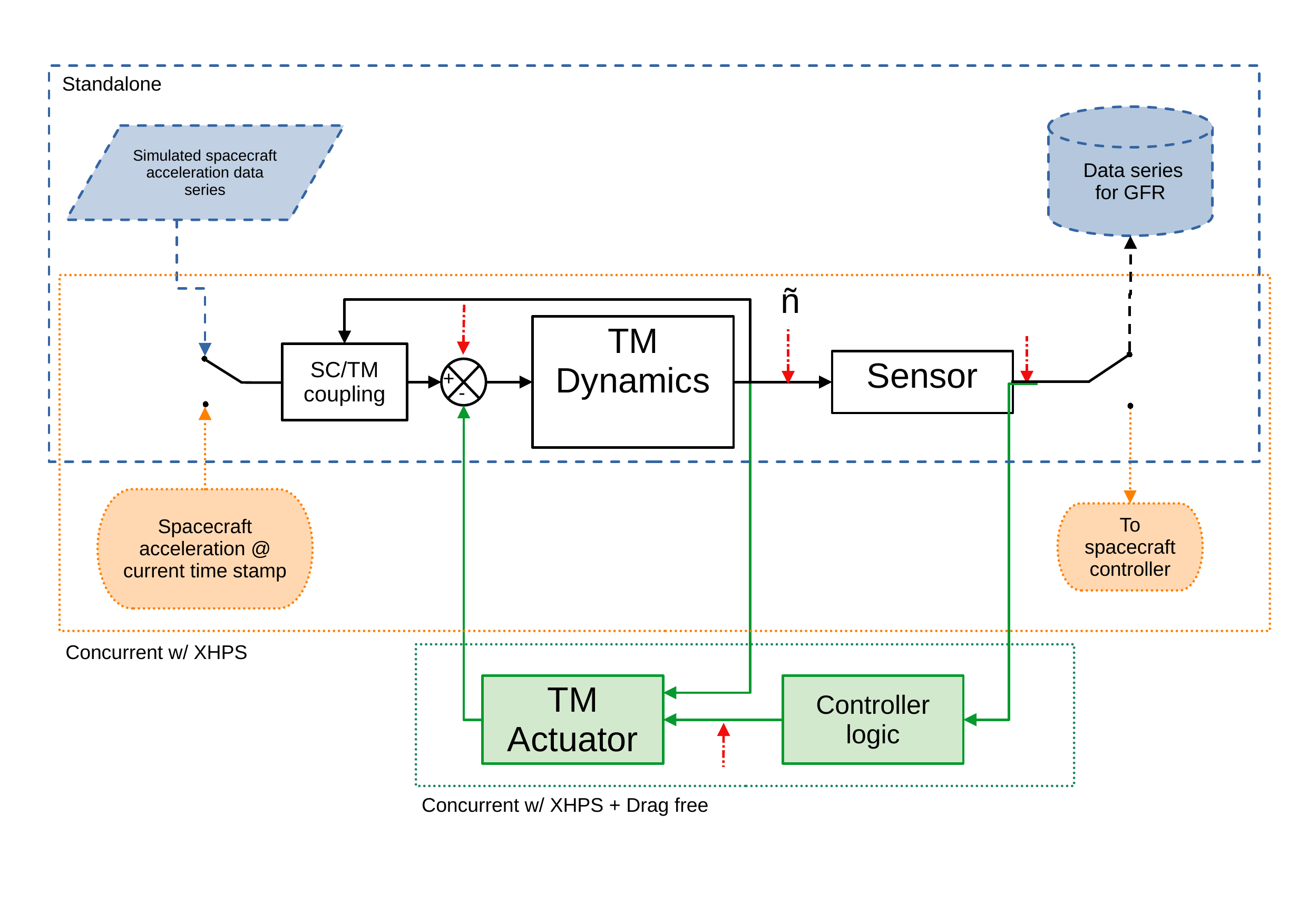}
\caption{Generalized ACC block diagram: Blue (dashed) blocks and connectors indicate a standalone simulation. Orange (dotted) blocks and connectors indicate concurrent satellite dynamics simulation with XHPS. Green blocks and connectors are only present in the non-drag free mode. Red arrows are points where a noise model $\tilde{n}$ can be added.}
\label{fig:acc_block_diagram}
\end{figure}

A generalized version of an ACC is represented in the block diagram in Fig.~\ref{fig:acc_block_diagram}. Although similar representations have been shown before \citep[][]{speake1997, Touboul1999, Frommknecht2003}, here we present a diagram that includes two different simulation types (standalone or with XHPS), mission scenarios (drag free and non-drag-free modes) and that is non-specific regarding its sensor technology. The core of the diagram is the sum of forces that affect the TM and the sensor model.
The orange (dotted) blocks and connectors indicate that the simulation is concurrent with the orbital dynamics simulation: the control system of the simulated satellite in XHPS uses the measured noisy accelerations in its control loop. The blue (dashed) lines indicate the standalone version. In this case, a pre-generated timeseries or ASD of non-gravitational forces for a given orbit and satellite are entered and a time or frequency response is simulated. Red arrows are points where an appropriate noise model $\tilde{n}$ can be added.
The green blocks and connectors are present in the non-drag compensated mode and are excluded in drag-free scenarios.

Although the noise models presented so far can already reproduce proposed models of instruments with good fidelity, it is not complete and we intend to add and refine noise sources.
In the simulation results that are presented, the dominant noise in the low frequency band is the actuation noise. This, however, might change in future works, when a more detailed model of thermal noise is implemented, since it is a component that increases towards lower frequencies and is simplified for the time being.

\section{Gradiometer modeling}
\label{sec: Gradiometers modelling}

\begin{figure}
  \centering
  \includegraphics[width=0.5\textwidth]{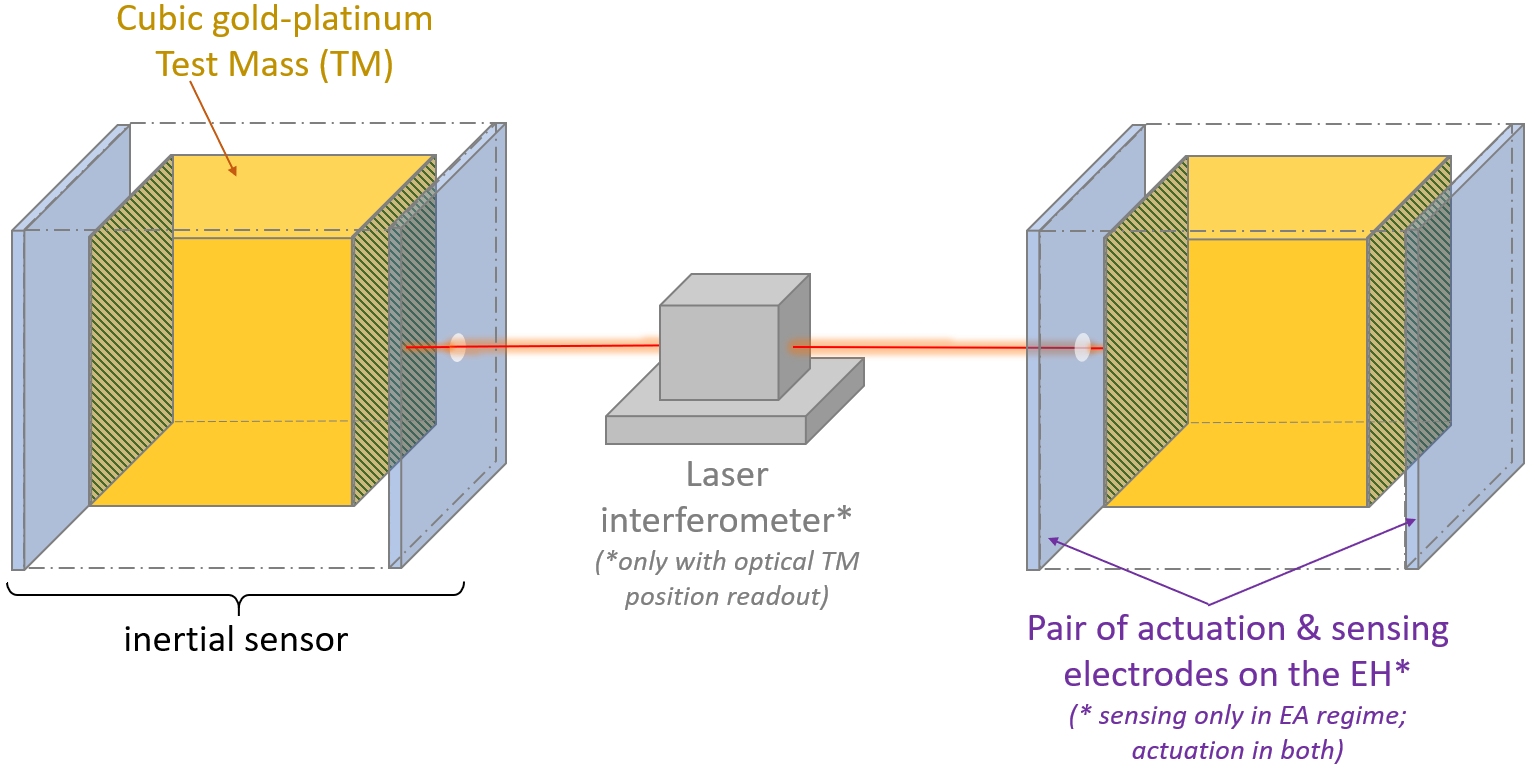}
  \caption{Scheme of the 1 degree of freedom optical gradiometer}
  \label{optical_gradio}
\end{figure}

The gradiometer mission GOCE measured Earth's static gravitational field with unprecedented precision so far \citep[][]{vanderMeijde.2015,Siemes.2019}. However, technology flown in LPF encourages the implementation of an optical metrology system which could be able to quantify not only spatial but also temporal variations of the gravity field with increased resolution. The optical gradiometer independently senses the position of two TMs by laser interferometry (see Fig.~\ref{optical_gradio}).

During the modeling of gradiometers, either electrostatic or optical, one has to consider the dynamics with care. Since the TMs are not co-located with the CoM of the spacecraft anymore, as it is in GRACE-like accelerometry, additional terms have to be taken into account, e.g., Euler, centrifugal and Coriolis accelerations \citep{Theil_S.}.

An equation giving the measurement of a perfectly-calibrated ACC was given by \citet{Siemes.2017}:

\begin{linenomath}\begin{equation}
    \vec{a}_i=-\left(\vec{V}-\vec{\Omega}^2-\dot{\vec{\Omega}} \right)\vec{r}_i+\vec{d}\text{,}
        \label{eqn: calibrated ACC}
\end{equation}\end{linenomath}

where $i=1,...,6$ -- ACC index (corresponding to six 3-axis accelerometers at GOCE); $\vec{V}$ -- gravity gradient; $\vec{\Omega}^2\vec{r}_i$ -- centrifugal acceleration; $\dot{\vec{\Omega}}\vec{r}_i$ -- angular acceleration of the satellite; $\vec{d}$ -- non-gravitational acceleration of the satellite and $\vec{r}_i$ -- is the vector from the satellite CoM to the \textit{i-th} ACC. Moreover,   $\vec{\Omega}^2$ and $\dot{\vec{\Omega}}$ represented in matrix forms are

\begin{linenomath}\begin{equation}
    \vec{\Omega}^2 = \begin{bmatrix} -\omega^2_y-\omega^2_z & \omega_x\omega_y & \omega_x\omega_z\\
    \omega_x\omega_y & -\omega^2_x-\omega^2_z & \omega_y\omega_z\\
    \omega_x\omega_z & \omega_y\omega_z & -\omega^2_x-\omega^2_y
    \end{bmatrix}\text{,}
\end{equation}\end{linenomath}

\begin{linenomath}\begin{equation}
    \dot{\vec{\Omega}} = \begin{bmatrix} 0 & -\dot{\omega}_z & \dot{\omega}_y\\
    \dot{\omega}_z & 0 & -\dot{\omega}_x\\
    -\dot{\omega}_y & \dot{\omega}_x & 0
    \end{bmatrix}\text{.}
\end{equation}\end{linenomath}

Differential mode accelerations are obtained by
\begin{linenomath}\begin{equation}
    a_{dij}=\frac{(\vec{a_i}-\vec{a_j})}{2}=-\frac{1}{2}\left(\vec{V}-\vec{\Omega^2}-\vec{\dot{\Omega}} \right) (\vec{r_i}-\vec{r_j}) \text{.}
        \label{eqn: DMA}
\end{equation}\end{linenomath}

After rearranging Eq. \eqref{eqn: DMA}, the gravity gradient in along-track direction is given as
\begin{linenomath}\begin{equation}
    V_{xx}=\frac{-2a_{d14x}}{L_x}-\omega_y^2-\omega_z^2 \text{,}
        \label{eqn: V_xx}
\end{equation}\end{linenomath}
where $\omega_y$ and $\omega_z$ are the angular rates of the satellite about the cross-track and nadir axes. For the remaining gravity gradient components, the reader is referred to \citet{Siemes.2017}.

The gradiometer GFR software GRADIO in Fortran takes into account all above-mentioned effects regarding TM dynamics in the case where ACC(s) are not co-located with the CoM of the satellite. By defining the gradiometer baseline between the ACCs and selecting the desired axis, i.e., along-track, cross-track or nadir, one can define specific gradiometer missions. The gradiometry results presented in this work utilize a GOCE-like cross-track gradiometer configuration with a baseline of \qty{0.5}{m} using a low Earth orbit with an altitude of about \qty{250}{\km} (similar to GOCE).

\section{Gravity field recovery}
\label{sec:GFR}

Gravity field solutions are computed to quantify the improvements from new types of instruments and concepts, e.g., novel ACCs, gradiometers as well as advanced satellite constellations, such as a Bender constellation or a combination of ll-SST with cross-track gradiometry.

\subsection{GFR with classical and novel sensors}

This section presents the comparison of the recovered gravity field models for a single satellite pair on a polar orbit, e.g., GRACE-like, using different inter-satellite range instruments and accelerometers. All amplitude spectral densities of different types of ACCs that are used in GFR here result from the modeling done in ACME. 

\begin{figure}
  \centering
  \includegraphics[width=0.5\textwidth]{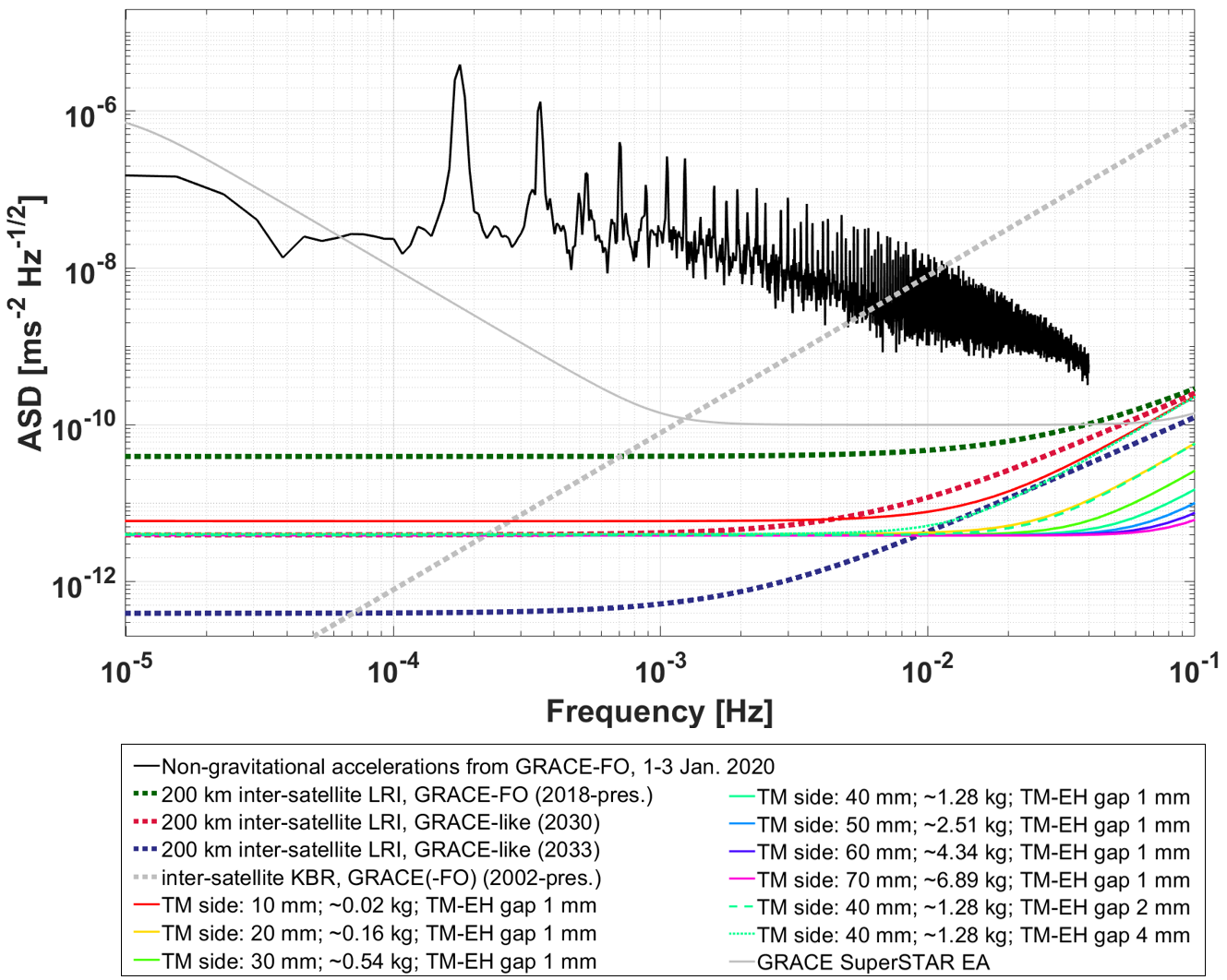}
  \caption{ASDs of SGRS-like EAs with varied parameter values compared to non-gravitational accelerations and inter-satellite LRI and KBR errors}
  \label{EA_SGRS_450km_parametrization}
\end{figure}

Figure \ref{EA_SGRS_450km_parametrization} shows the amplitude spectral densities of the  SGRS-like electrostatic ACCs with variations of different parameters, i.e., the dimensions of the TM and the size of the gap between the TM and the electrode housing (TM-EH), as well as non-gravitational accelerations and inter-satellite Laser Range Interferometer and K-band ranging errors. The parameters corresponding to the continuous light green line ($\qty{30}{mm}$ side length, $\qty{0.54}{kg}$ mass and $\qty{1}{mm}$ gap) are equal to the model proposed in \citet{Alvarez2022} and the resulting sensitivity curve, independently calculated with our tool, is similar to the one presented in Fig.~13 of their work.

The ASD of the LRI noise for the present-day GRACE-FO mission (green dotted line in Fig.~\ref{EA_SGRS_450km_parametrization}) was assumed in this study as
\begin{linenomath}\begin{multline}
     \sasdf =    \left[\frac{L}{\qty{1}{m}}\cdot10^{-15}\sqrt{\frac{\qty{1}{\Hz}}{f}} + 10^{-12} \left(\frac{\qty{1}{\Hz}}{f}\right)^2\right] \\
      \times \left(\twopif\right)^2 \mpsssqrthzfrac{} \text{,} \label{eqn:LRI_GRACE-FO 2022}
\end{multline}\end{linenomath}
where $L$ is the inter-satellite distance, for which a value of \qty{200}{\km} is used in the simulations. The first term in the bracket proportional to $L$ accounts for the cavity intrinsic thermal noise \citep[][]{Francis.2015}, while the second term approximates the low-frequency thermal stability of the cavity.

The expected behavior of the ASD of the inter-satellite LRI error for future GRACE-like gravimetry missions in 2030 (red dotted line in Fig.~\ref{EA_SGRS_450km_parametrization}) is assumed as
\begin{linenomath}\begin{multline}
     \sasdf =  
     \left[\frac{L}{\qty{1}{m}}\cdot10^{-15}\sqrt{\frac{\qty{1}{\Hz}}{f}} + 10^{-13} \left(\frac{\qty{1}{\Hz}}{f}\right)^2\right]\\
     \times\left(\twopif\right)^2 \mpsssqrthzfrac{} \text{.}\label{eqn:LRI_GRACE-FO 2030}
\end{multline}\end{linenomath}
For GRACE-like gravimetry missions beyond 2033, depicted by the blue dotted line in Fig.~\ref{EA_SGRS_450km_parametrization}, it is
\begin{linenomath}\begin{multline}
     \sasdf = \left[\frac{L}{\qty{1}{m}} \cdot 5\times10^{-16}\sqrt{\frac{\qty{1}{\Hz}}{f}} \right.
       \left. + 10^{-14}\left(\frac{\qty{1}{\Hz}}{f}\right)^2\right]\\
       \times\left(\twopif\right)^2 \mpsssqrthzfrac{} \text{.}\label{eqn:LRI_GRACE-FO 2033}
\end{multline}\end{linenomath}

\begin{figure*}[h!]
  \centering
  \includegraphics[width=1\textwidth]{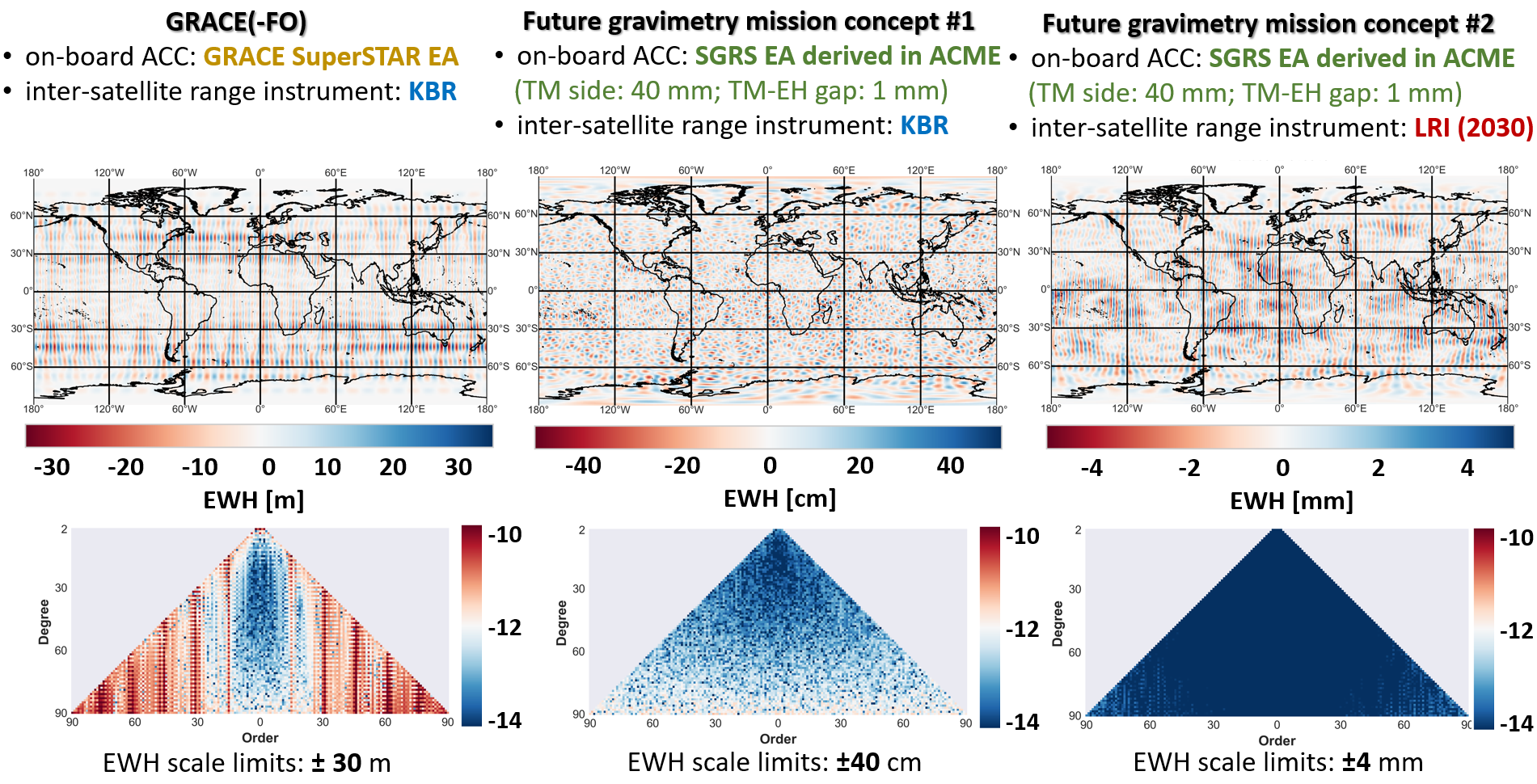}
  \caption{Comparison of recovered gravity fields (raw data, without post-processing and filtering) between simulated GRACE-FO and future gravimetry mission concepts w.r.t. EIGEN-6c4 in terms of EWH up to spherical harmonic degree 90}
  \label{GFR_NGGM}
\end{figure*}

The first column in Fig.~\ref{GFR_NGGM} represents GFR simulations based on the current gravimetry mission GRACE-FO with an EA and KBR. The classical EA ASD noise curve was introduced in Eq. \eqref{eqn:ASD superSTAR EA [Daras2017]}.

The ASD of the KBR sensor (grey dotted line in Fig.~\ref{EA_SGRS_450km_parametrization}) from \citet{frommknecht2006integrated} can be represented as
\begin{linenomath}\begin{equation}
\label{eqn:equation ASD KBR}
  \sasdf =   \left(\twopif\right)^2 \cdot 2\times 10^{-6} \mpsssqrthzfrac{}\text{.}
\end{equation}\end{linenomath}

Here, one month of observations without errors due to background models, i.e., temporal aliasing errors, were considered. An orbit with an altitude of \qty{450}{\km}, an inclination of \qty{89}{\degree} and an inter-satellite separation of \qty{190}{\km} was chosen. The estimated error degree variances of the unitless SH coefficients are shown in terms of EWH in a geographic representation. A characteristic feature of the recovered gravity field from this mission is the North-South striping effect that comes due to a variety of reasons, such as the drift of the EAs in low frequencies or technical properties of the capacitive type of sensors, as well as orbital resonance effects \citep[][]{Kvas.2019}. A similar striping behavior was also demonstrated by \citet[][]{Knabe.2022} in the comparison of an EA with a hybrid ACC. The amplitudes of the difference to the reference gravity field named European Improved Gravity model of the Earth by New techniques (EIGEN)-6c4 corresponding to the `GRACE' scenario with EA and KBR instruments are \qty{\pm30}{\m} EWH. The results were generated without applying any post-processing and filtering. 

The bottom left panel in Fig.~\ref{GFR_NGGM} shows the two dimensional SH error spectrum, i.e., the relative error of each spherical harmonic coefficient, given as the difference between the estimated gravity field and the reference model EIGEN-6c4. One can see the pronounced stripes occurring every 16 orders which is caused by the orbital frequency and the resulting repeat patterns of the groundtracks \citep[][]{Seo.2008,Cheng.2017}.

Analogously, the middle column of Fig.~\ref{GFR_NGGM} shows the result for the future gravimetry mission concept \#1 when replacing the EA named Super Space Three-axis Accelerometer for Research (SuperSTAR) by a SGRS-like EA.
The difference to the reference gravity field EIGEN-6c4 now has smaller amplitudes of \qty{\pm40}{\cm} EWH in comparison to the GRACE-FO case (amplitudes \qty{\pm30}{\m}). In this solution, the following sensors were considered: KBR for the inter-satellite range measurement (same as for the GRACE-FO case) and an SGRS-like EA with a cubic gold-platinum TM with a side length of \qty{40}{\mm} and a TM-EH gap of \qty{1}{\mm}. The ASD of this scenario is given as

\begin{linenomath}\begin{multline}
   \label{eqn:equation EA SGRS from ACME (TM side: 40mm; gap 1 mm)}
  \sasdf = 10^{-10} \mpsssqrthz{} ~\\
  \times\left[14.35 \left(\frac{f}{\qty{1}{Hz}} \right)^2 - \frac{0.1655 f}{\qty{1}{Hz}}  + 3.81\times10^{-2} \right].
\end{multline}\end{linenomath}

The ASD equation as 2nd order polynomial which was fitted to the frequency response of the linearized SGRS model in ACME. For that, a typical spectrum of non-gravitational accelerations was given as the input.

In addition to the overall decrease in amplitudes, the North-South striping effect is also significantly reduced. The improvement in the accuracy of the estimated gravity field can also be seen in the fact that the difference to the reference gravity field used for the synthesis is smaller. Lower the numbers in the color bar indicate better matches between the recovered gravity field and the reference gravity field.

The right column in Fig.~\ref{GFR_NGGM} corresponds to the future gravimetry mission concept \#2 using `novel' sensors for both inter-satellite ranging and accelerometery, i.e., an LRI expected for GRACE-like missions in 2030 (Eq. \eqref{eqn:LRI_GRACE-FO 2030}) and an SGRS-like EA (same as in the previous case). The EWH amplitudes of the difference to the reference gravity field EIGEN-6c4 correspond to \qty{\pm4}{\mm}.

The SH two-dimensional spectrum (bottom right panel) also shows a significant improvement in comparison to the other GFR simulations. In other words, the recovered gravity field from a GRACE-like mission with an optical ACC and an LRI on board is much closer to the reference gravity field EIGEN-6c4.

\begin{figure}
\centering
\includegraphics[width=.5\textwidth]{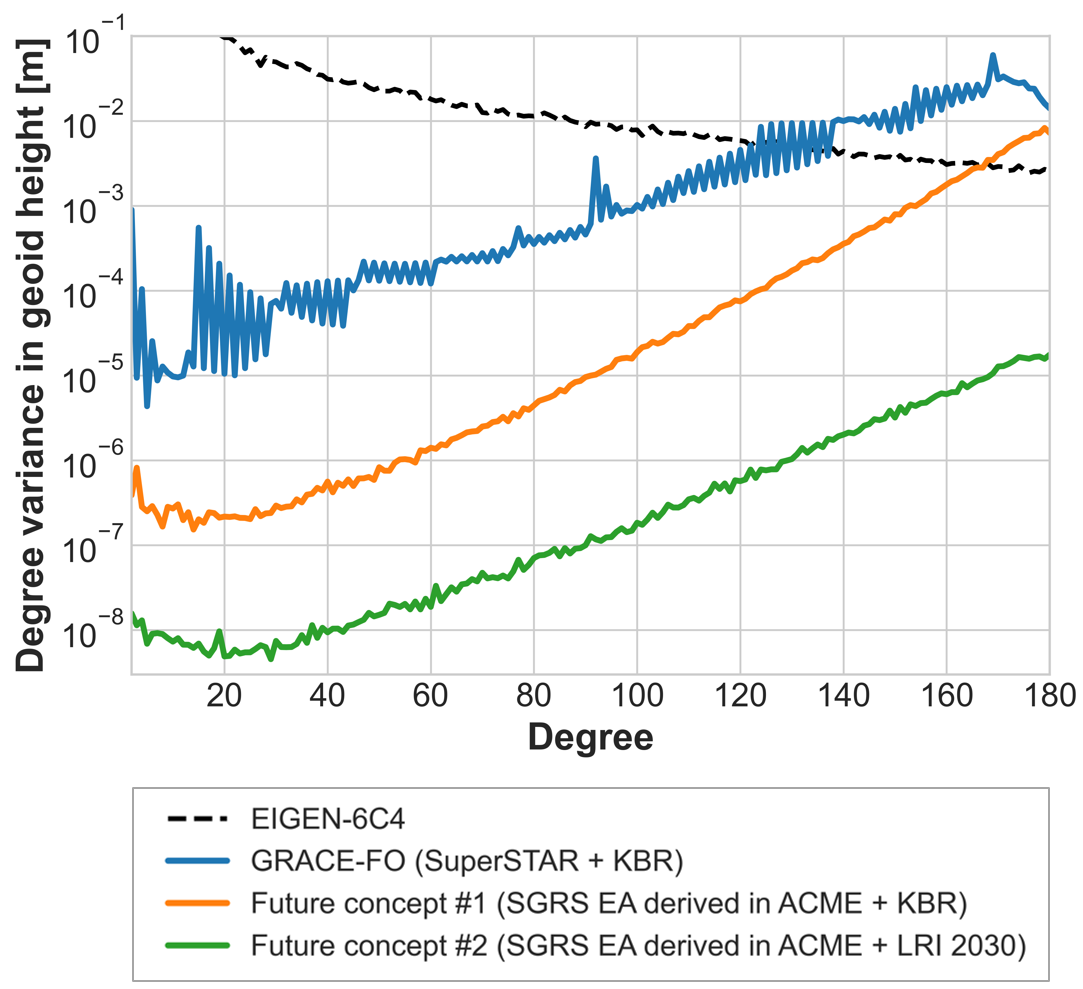}
\caption{Averaged error degree variance per specific degree of the GRACE-FO and future gravimetry mission concepts in terms of geoid height w.r.t. EIGEN-6c4 up to spherical harmonic degree 180}
\label{GFR_NGGM_degree_variances}
\end{figure}

Figure \ref{GFR_NGGM_degree_variances} represents the corresponding averaged error degree variance per specific degree of the simulated GRACE-FO and future gravimetry mission concepts in terms of geoid height. The GRACE-FO mission could resolve the static gravity field up to degree 130 with the maximum level of accuracy  \qty{10^{-4}}{\m}, while the future gravimetry mission concepts \#1 and \#2 accurate up to \qty{10^{-7}}{\m} and \qty{10^{-8}}{\m}. Also, the curves from the future concepts are much smoother than the one from the GRACE-FO mission. This is consistent with the results shown in Fig.~\ref{GFR_NGGM}. Thus, the North-South striping problem due to the orbital resonance effects and drift of the EA in the low frequency domain could be reduced with novel sensors.

\subsection{GFR from Bender constellation}
\label{subsec GFR Bender}
One of the goals of future gravimetry missions is to improve the understanding of mass transport in the Earth system. Long repeat periods of the orbit for GRACE(-FO) ($\sim$ 30 days) and GOCE ($\sim$ 60 days) cause an inhomogeneous distribution of the groundtracks and consequently a lack of signal. Additionally, the peculiar North-South striping effect in the recovered gravity field solutions of GRACE(-FO) leads to a degradation of the results and a significant loss of the signal.

\begin{figure}[h]%
\centering
	    {\includegraphics[width=0.5\textwidth]{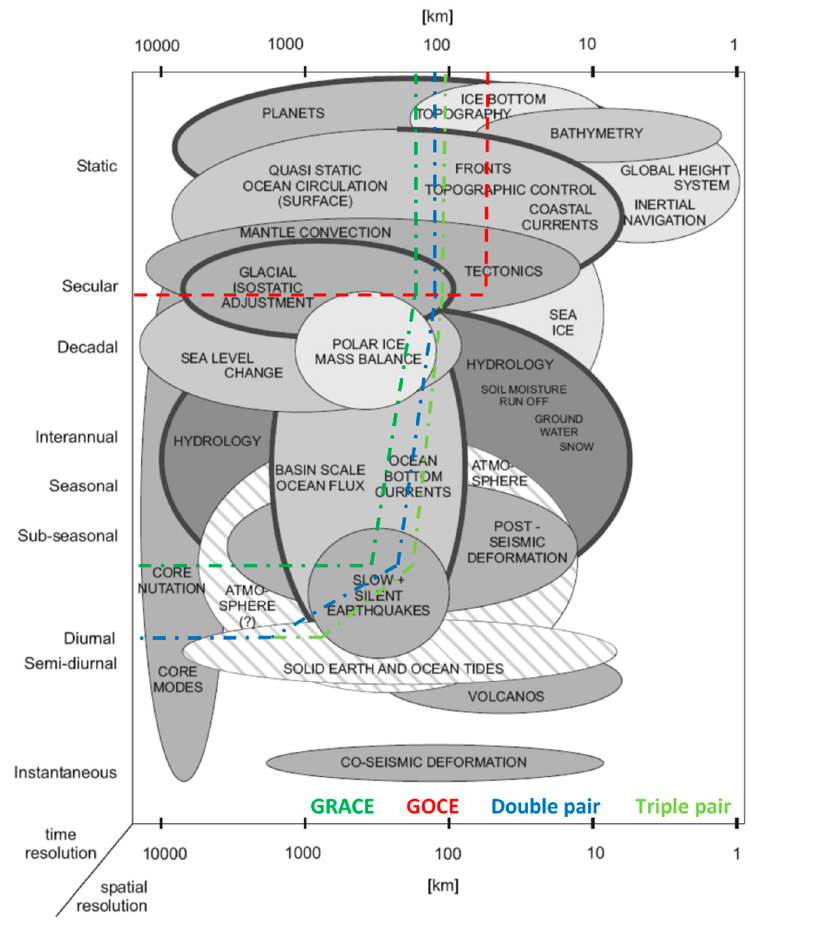}%
			}%
			\hfil%
\caption{Scheme of the temporal and spatial components of the gravity field signal covered by GRACE, GOCE and two future gravimetry mission concepts: double-pair Bender constellation and triple-pair constellation \citep{Purkhauser2020}} %
\label{triple_Purkhauser}%
\end{figure}

Most of these problems could be solved by a so-called Bender constellation. This constellation consists of two GRACE-like satellite pairs, one on a near-polar and one in an inclined orbit. \cite{Purkhauser2020} presented a scheme of the temporal and spatial components of the gravity field covered by existing satellite gravimetry missions, i.e., GRACE and GOCE, as well as several future gravimetry mission concepts like a Bender-type constellation or a triple pair, see Fig.~\ref{triple_Purkhauser}. The Bender-type constellation could be an optimal trade off through the costs of the mission and the technical difficulties in controlling the motion of the satellites in comparison to the triple-pair constellation. Moreover, the Bender double-pair constellation could enhance both spatial and temporal resolutions thanks to the much denser coverage of the globe, which allows to self-dealias high-frequency atmospheric and oceanic signals and even provide gravity solutions up to diurnal timescales \citep{Purkhauser2020}. 

The selection of the inclination parameter for the inclined orbit depends on the applications and the area of interest for the gravity field recovery. Typically, the inclination parameter lies in the range $\qty{63}{\degree}\leqslant i \leqslant\qty{75}{\degree}$. Different authors and research groups claimed various inclination parameters from the above-mentioned range with different argumentation. For example, \cite{Cesare2010} and \cite{Elsaka2014} proposed $i=\qty{63}{\degree}$ as the optimum for the double-pair configuration to achieve a uniform and fast coverage on the global scale. A group of scientists from the Technical University of Munich \citep{Purkhauser2019} selected $i=\qty{70}{\degree}$ for their analysis and \cite{Wiese2012} chose $i=\qty{72}{\degree}$. If the primary goal of a gravimetry mission is the observation and determination of the continental hydrology, then lower inclinations would be preferable. On the other hand, if one is looking for the glaciers in the polar regions, higher inclinations are more suitable.

In our study, GFR simulations for the Bender-type constellation were run with an LRI inter-satellite range measurement system, where the noise ASD was based on \citet[][c.f. Fig.~5]{Abich2019}:
\begin{linenomath}\begin{multline}
    \label{eqn:equation ASD LRI [Abbich2019]}
    \sasdf =   
    \left[2\times 10^{-9} \cdot \left(\frac{\qty{1}{\Hz}}{f}\right)^{0.3}+\left(\frac{\qty{0.3}{\uHz}}{f}\right)^3 \right] \\
    \times\left(\twopif\right)^2 \mpsssqrthz ~ \\
\end{multline}\end{linenomath}
and for the accelerometer performance, the GRACE technical requirement by \citet[][]{Kim.2002, Darbeheshti2017} was taken:
\begin{linenomath}\begin{equation}
    \label{eqn:equation 'optimistic' EA}
    \sasdf = \sqrt{1+\frac{\qty{5}{\mHz}}{f}} \times  \mpsssqrthz{10^{-10}} \text{.}
\end{equation}\end{linenomath}

\begin{figure*}
  \centering
  \includegraphics[width=1\textwidth]{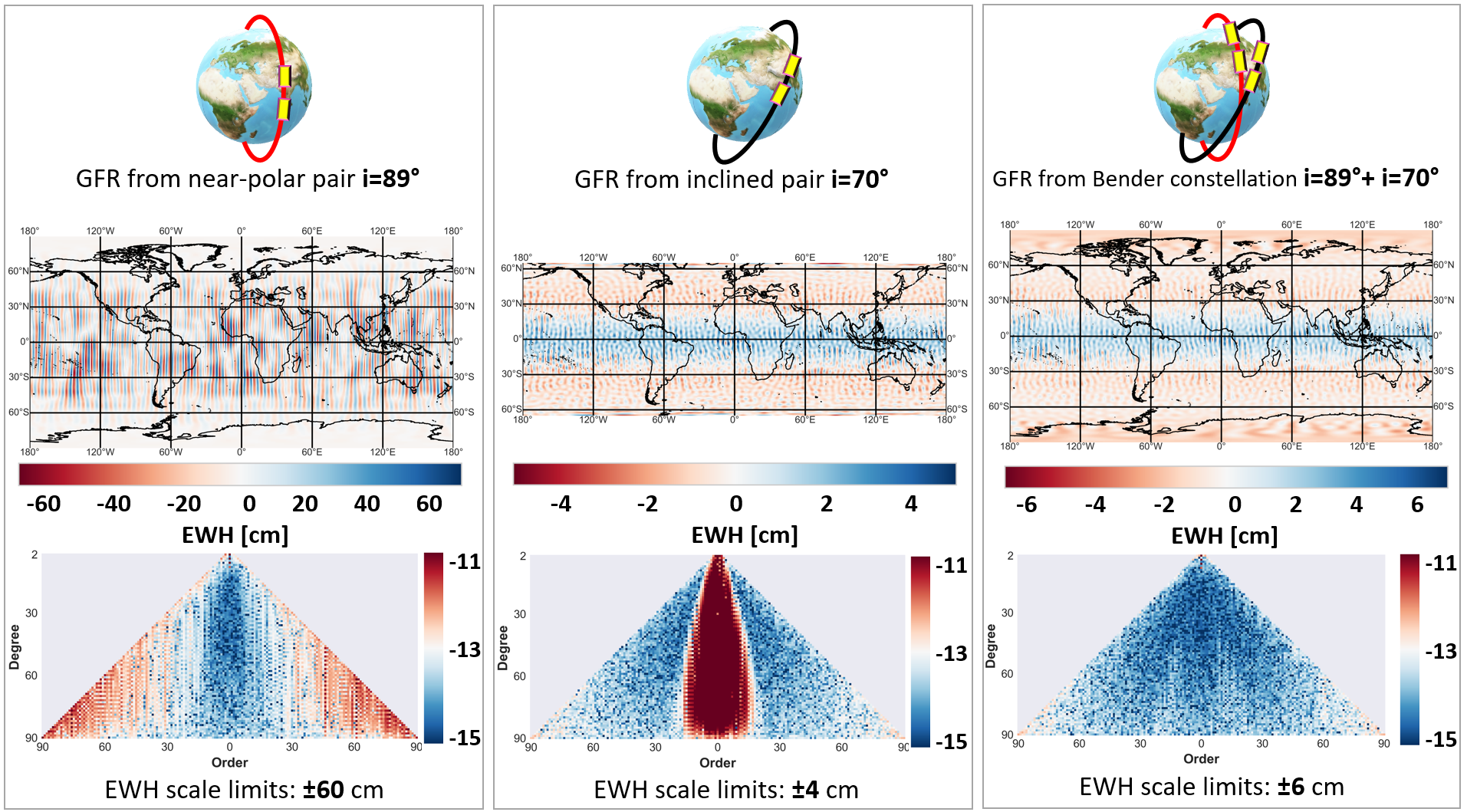}
  \caption{Recovered gravity field and spherical harmonic error spectra from the near-polar orbit and inclined orbit ($i=\qty{70}{\degree}$) with LRI and GRACE tech. requirement EA w.r.t. EGM2008 gravity field model}
  \label{GFR_Bender}
\end{figure*}

The left column in Fig.~\ref{GFR_Bender} shows the estimated gravity field from the near-polar orbit with LRI and an EA following the GRACE technical requirement (errors simulated using Eq. \eqref{eqn:equation ASD LRI [Abbich2019]} and Eq. \eqref{eqn:equation 'optimistic' EA}) w.r.t. the Earth Gravitational Model (EGM) 2008 reference gravity field model. EWH amplitudes of the difference to the reference gravity field are \qty{\pm0.6}{\m}. The lower panel shows the corresponding SH error spectrum for the recovered gravity field.

Analogously, the results for the inclined satellite pair with an orbit inclination of $i=\qty{70}{\degree}$ are depicted in the middle part of Fig.~\ref{GFR_Bender}. The signal amplitudes of the recovered gravity field differ by $\qty{\pm 0.04}{\m}$ EWH in comparison to the EGM2008 reference gravity model, due to the denser measurements and smaller variance of the results. In the SH error spectrum, one can see the wedge corresponding to the less accurate zonal and near-zonal SH coefficients as the polar regions are not covered.  

Finally, the right column in Fig.~\ref{GFR_Bender} corresponds to the combination of the GFR from the near-polar and inclined satellite pairs. The combination is performed at the level of normal equations, with the a posteriori variances from two orbits, acting as a weighting factors, to obtain the combined solution. Since the residuals for the inclined satellite pair were smaller than those for the near-polar satellite pair, the combined gravity field solution has also low residuals. Therefore, the recovered gravity field from the inclined orbits gets a larger weight than that from the near-polar satellite pair. The SH error spectrum of the combined solution does not show the wedge that is visible in the corresponding SH error spectrum of the inclined satellite pair, and the tesseral SH coefficients are also determined more accurately in the combined solution than in the near-polar solution. Thus, the difference to the reference gravity field EGM2008 from the combined solution has a low amplitude and is homogeneously distributed over the entire spectrum up to degree and order 90.

\subsection{GFR from different gradiometer concepts} 

As GOCE was the only gradiometer satellite mission up to now, the gradiometry comparison (in cross-track direction $V_{yy}$) will be performed w.r.t. GOCE's high-sensitive axis gradiometer, using the GOCE accelerometer noise model from \citet{Touboul.2016} and \citet{Marque2010}:

\begin{linenomath}\begin{multline}
    \label{eqn:GOCE high-sen axis}
     \sasdf =  2\times10^{-12} \mpsssqrthz ~ \\
     \times\sqrt{\left(\frac{\qty{1}{\mHz}}{f}\right)^2+1+\left( \frac{f}{\qty{0.1}{\Hz}}\right)^4} \text{.}
\end{multline}\end{linenomath}

\begin{figure}
  \centering
  \includegraphics[width=0.5\textwidth]{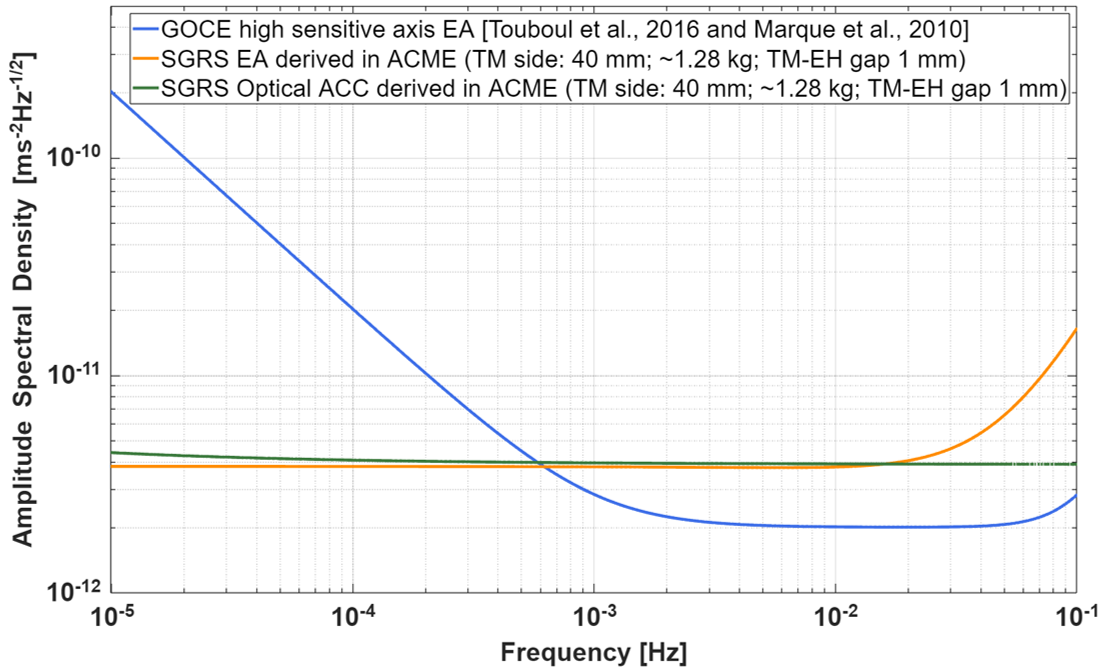}
  \caption{ASD of the accelerometers that form the gradiometers}
  \label{GFR_gradio_comparison_1}
\end{figure}

The gradiometry results presented in this section were all computed for the same scenario, which is a simulated near-polar orbit with inclination $i=\qty{89}{\degree}$ and altitude $h=\qty{246}{\km}$ for 1 month and with a cross-track baseline $b=\qty{0.5}{m}$. No errors due to background modeling were considered. Fig.~\ref{GFR_gradio_comparison_1} shows the ASD of the ACCs that form the gradiometers. Here, SGRS-like electrostatic and optical accelerometers were modeled in ACME and compared with the high-sensitive axis gradiometer of GOCE. Both simulated ACCs have the same parameters, i.e., a cubic gold-platinum TM with side length \qty{40}{\mm}, weight \qty{1.28}{\kg} and TM-EH gap of \qty{1}{\mm}. Both simulated SGRS derived in ACME with nearly white noise behavior at low frequencies are likely too optimistic, so an updated version of the simulator will re-examine the noise at these frequencies in further studies

\begin{figure}
  \centering
  \includegraphics[width=0.5\textwidth]{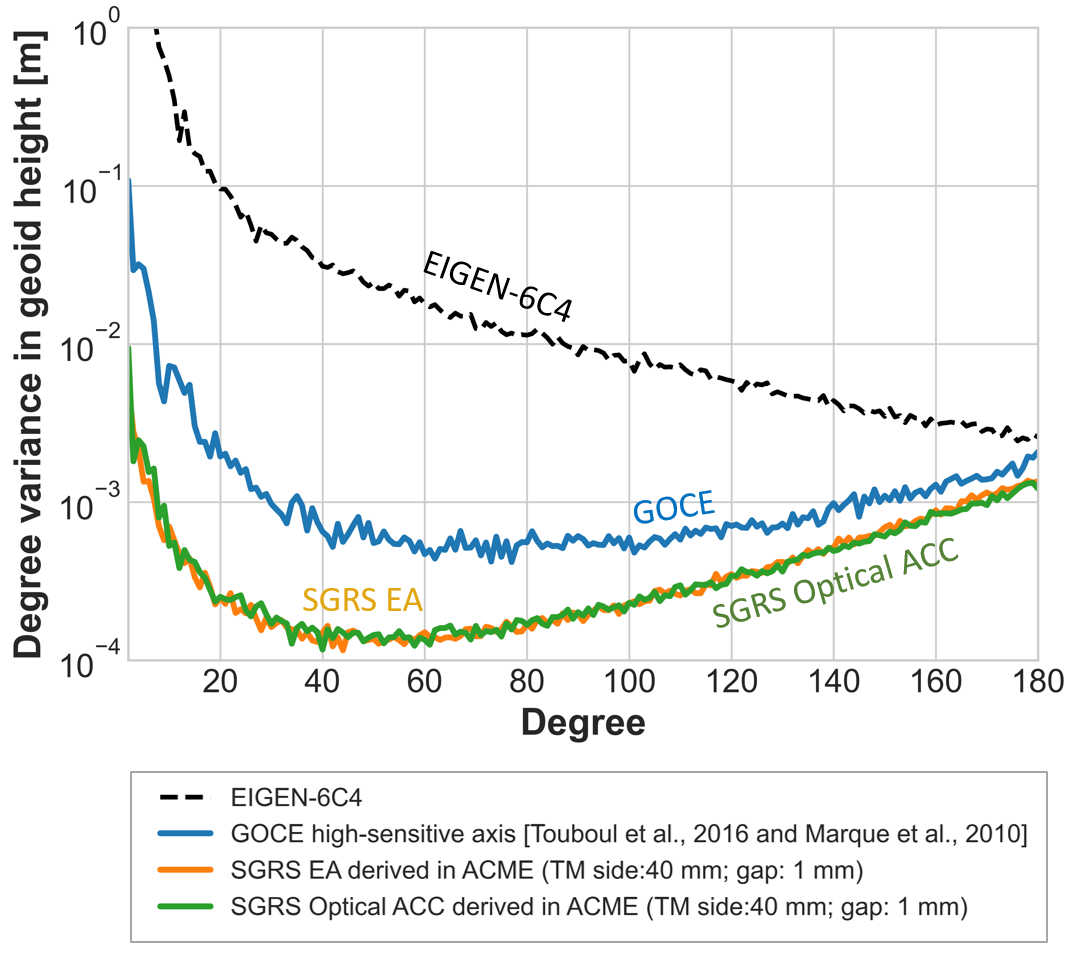}
  \caption{Averaged degree variance per specific degree in geoid height (m) for the cross-track component $V_{yy}$ for different gradiometer models}
  \label{GFR_gradio_comparison_3}
\end{figure}

Fig. \ref{GFR_gradio_comparison_3} represents the mean error degree variance of the geoid height per degree from the cross-track component $V_{yy}$. A significant improvement between GOCE and the novel SGRS is noticeable over all degrees. The difference between the electrostatic and optical SGRS-like accelerometers is not that significant, since the difference in their ASDs only appears above \qty{0.01}{\hertz} and is therefore only relevant for very high degrees. 

\begin{figure}
  \centering
  \includegraphics[width=0.5\textwidth]{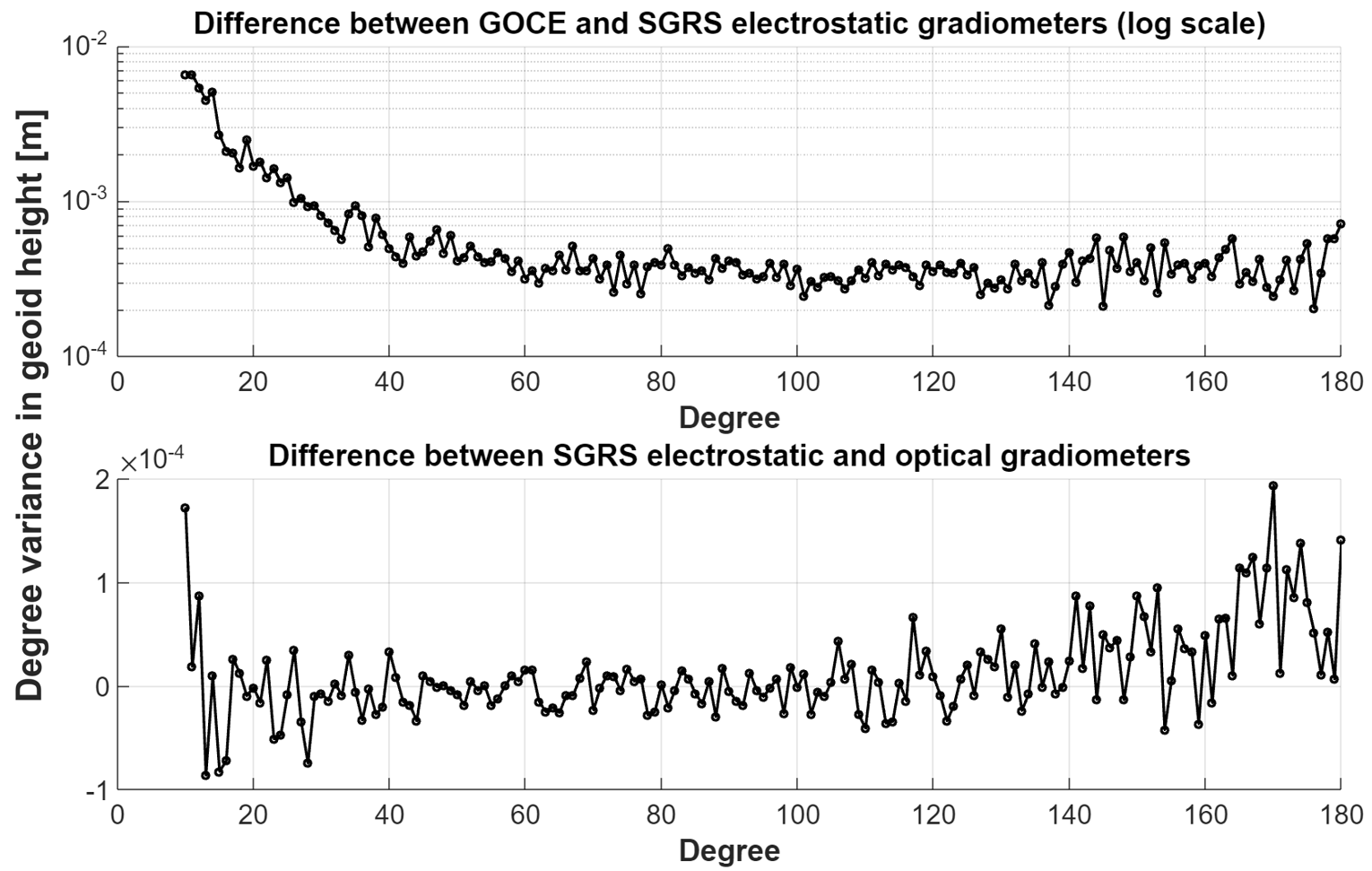}
  \caption{Degree variance in geoid height (m) of the difference between various gradiometer solutions from SH degree 8}
  \label{GFR_gradio_comparison_2}
\end{figure}

Fig. \ref{GFR_gradio_comparison_2} represents the difference between the solutions of the GOCE and the SGRS EA gradiometers in logarithmic scale (upper graph) and the difference between the solutions of EA and SGRS optical gradiometers in linear scale (lower graph). The curves start at SH degree 8, as the normal gravity field signal was not included in the GFR simulations. There is a large difference in the upper graph that is decreasing towards degree 60, while there is an inverse trend in the bottom graph. The difference between SGRS electrostatic and optical accelerometers is due to the expected improved performance of the optical SGRS.

\subsection{GFR from the combination of ll-SST and cross-track gradiometry} 

The North-South striping effect in gravity field solutions from the current gravimetry mission GRACE-FO and upcoming satellites in orbits with high inclination (i= \qty{89}{\degree}) can be overcome by novel sensors and measurement techniques. Moreover, GRACE-like missions are able to detect temporally changing gravity signals down to sub-seasonal level, while the GOCE-like missions mostly resolve static gravity field signals down to one hundred kilometers. The combination of ll-SST and cross-track gradiometry in one mission could benefit from the advantages of the two techniques.

\begin{figure}[h]%
\centering
	    {\includegraphics[width=0.48\textwidth]{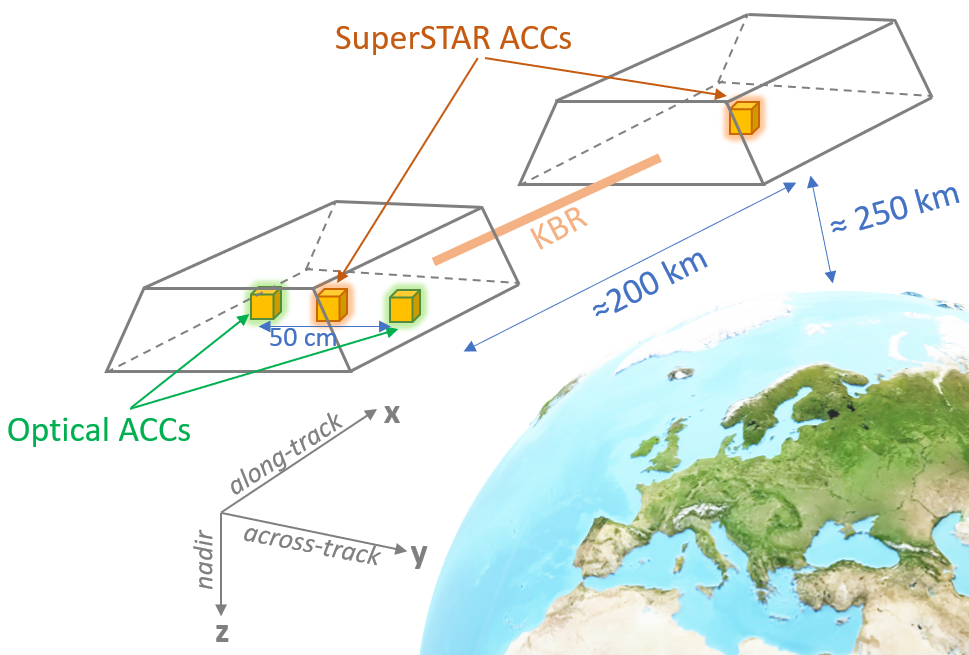}%
			}%
			\hfil%
\caption{Scheme of the combination of the low-low satellite-to-satellite tracking and cross-track gradiometry in a possible future gravimetry mission} %
\label{NGGM_combination_scheme}%
\end{figure}

Here, the idea of combining ll-SST and gradiometry is the utilization of a polar GRACE-like satellite pair with a \qty{246}{\km{}} altitude while placing a gradiometer on one of the spacecrafts in cross-track direction. This gradiometer consists of a pair of ACCs symmetrically distanced from the central ACC that is co-located with the CoM of the satellite (see Fig.~\ref{NGGM_combination_scheme}). The two satellites act as a big gradiometer with a \qty{200}{\km{}} baseline. For the ll-SST classical instruments were considered (SuperSTAR electrostatic ACCs and KBR) and for the cross-track gradiometer novel optical SGRS-like ACCs derived in ACME. The parameters of the optical ACCs are: cubic gold-platinum TM with a side length of \qty{40}{\mm} and a TM-EH gap of \qty{1}{\mm}.

\begin{figure*}[hbt!]%
\centering
{\includegraphics[width=1\textwidth]{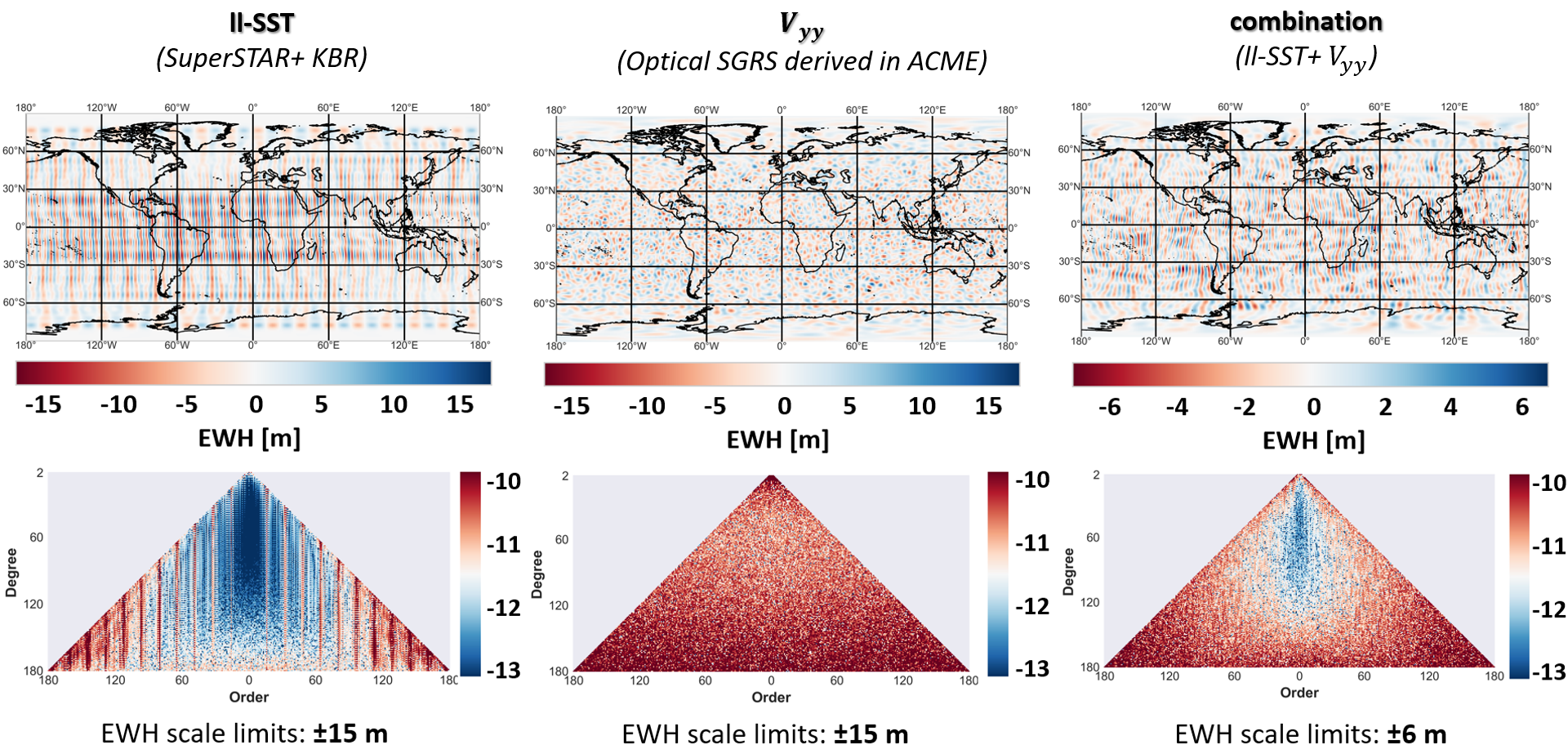}%
			}%
\caption{Recovered gravity fields and SH error spectra from ll-SST (SuperSTAR EA and KBR), cross-track gradiometry (SGRS optical ACC derived in ACME) and their combination w.r.t. EIGEN-6c4} %
\label{GFR_combination}%
\end{figure*}		

The gravity field was recovered from the combined observations based on the a posteriori variances of the individual solutions, similar to the GFR calculations of a Bender constellation, see section \ref{subsec GFR Bender}. 

\begin{figure}[h!]%
\centering
	    {\includegraphics[width=0.5\textwidth]{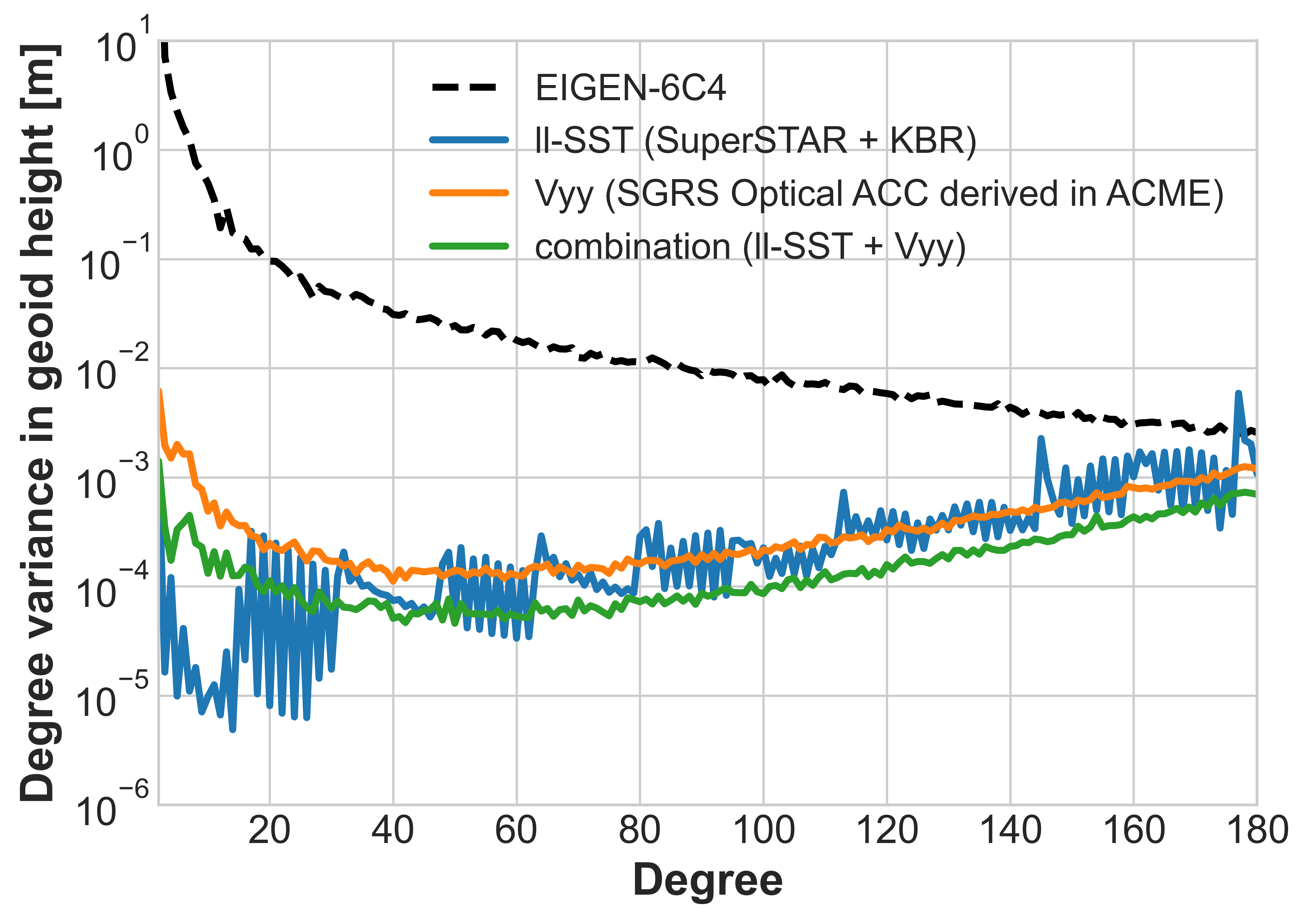}%
			}%
			\hfil%
\caption{Averaged variance per specific degree in geoid height from ll-SST (SuperSTAR EA and KBR), cross-track gradiometry (SGRS optical ACC derived in ACME) and their combination w.r.t. EIGEN-6c4} %
\label{error_degree_variances_combination}%
\end{figure}

Averaged values of the of the error degree variance in geoid height for the ll-SST, cross-track gradiometry and their combination up to SH degree 180 are shown in Fig.~\ref{error_degree_variances_combination}. Here, the assumptions were a measurement span of one month without errors due to background models, a drag-free regime and orbits with an altitude of \qty{246}{\km}, an inclination of \qty{89}{\degree} and an inter-satellite separation of \qty{193}{\km}. The blue curve in Fig.~\ref{error_degree_variances_combination} corresponds solely to the ll-SST solution. Periodic oscillations were already discussed earlier (see Fig.~\ref{GFR_NGGM_degree_variances}). They are caused by the drift of the SuperSTAR EA at low frequencies and orbital resonance effects \citep[][]{Kvas.2019}. The orange curve in Fig.~\ref{error_degree_variances_combination} represents the solution from the cross-track gradiometry. The combined solution (green curve) shows the improvement that can be obtained up to degree 180. 

In terms of the difference between recovered and reference gravity fields in EWH, Fig.~\ref{GFR_combination} shows the gravity field solutions on a global map and the SH error spectra. The left column corresponds to the ll-SST scenario, which resulted in a difference of \qty{\pm15}{\m} EWH. The middle column shows the cross-track gradiometry result $V_{yy}$ with differences to the reference gravity field at the same order of \qty{\pm15}{\m} EWH, but without stripes in North-South direction. Finally, the right column shows the combined result. The recovered gravity field model differs by only \qty{\pm6}{\m} EWH from EIGEN-6c4. The 2D SH error spectra (lower row) also shows the improvement of the combined solution.

\section{Conclusions and outlook}
\label{sec:conclusions}

We demonstrated the capability of modeling multiple types of novel concepts of gravimetry missions using different software modules and toolboxes. Starting from the detailed satellite orbital dynamics simulator (XHPS), that can consider various environment conditions, the simulation of current and future ACCs with different characteristics in ACME was discussed, and the generation of GFR solutions in QACC and GRADIO including diverse combination techniques was shown.  
    
In this paper, we validated our modeling approach, by implementing the parameters defined in concepts from other research groups \citep{Alvarez2022, Weber2022} in our software, and by demonstrating similar performance levels. It has been shown that novel ACCs will largely improve the gravity field recovery, due to the absence of drifts at low frequencies. It is worth noting that the simulated accelerometer models derived in ACME with almost white noise behavior at frequencies below $\qty{1}{\mHz}$ are likely too optimistic. The next version of the simulator will revisit the noise budget at the low frequency domain.

However, at that point, the inter-satellite ranging techniques will then potentially become the bottleneck. Even the expected accuracies of an LRI in 2030-2033 will limit the measurement accuracy, and also obscure the full potential of the novel ACCs. It is important to notice here that, in order to show the maximum benefit of the novel ACCs, the background models, which are a major limiting factor today, have not been considered in this study.

Using the novel ACCs in gradiometry could significantly improve the spatial resolution of the recovered gravity field compared to GOCE.

It has also been demonstrated that the combination of alternative satellite formations, e.g., double-pair Bender-type missions using novel sensors, could dramatically improve the spatial resolution of the recovered gravity field. The combination of ll-SST gravimetry and cross-track gradiometry would also provide better gravity field solutions.

Further research aims to streamline the workflow and add more features to the ACC modeling toolbox ACME. Then, the study will proceed evaluating additional combinations of mission scenarios, instrument payloads, and GFR techniques including the effects from the background models, so that novel concepts of future gravimetry missions can be quantified even more realistically.

\section{Acknowledgments}
\label{sec:Acknowledgments}
This work is funded by: Deutsche Forschungsgemeinschaft (DFG, German Research Foundation) – Project-ID 434617780 – SFB 1464.

%\vfill
%% Bibliography
%% Author year style
\bibliographystyle{jasr-model5-names}
\biboptions{authoryear}
\bibliography{refs}

\section{Appendix}
\label{sec:Appendix}
\begin{table}[H]
\caption {List of acronyms}
\begin{tabular}{p{1.9cm} p{6cm}}
\textbf{ACC} & Accelerometer \\
\textbf{ACME} & Accelerometer Modeling Environment   \\
\textbf{ASD} & Amplitude Spectral Densities    \\
\textbf{CoM} & Center of Mass   \\
\textbf{EA}  & Electrostatic Accelerometers  \\
\textbf{EH} & Electrode Housing \\
\textbf{EIGEN} & European Improved Gravity model of the Earth by New techniques  \\
\textbf{EGM} & Earth Gravitational Model \\
\textbf{ESA} & European Space Agency  \\
\textbf{EWH} & Equivalent Water Height \\
\textbf{GFR} & Gravity field recovery \\
\textbf{GOCE} & Gravity Field and Steady-State Ocean Circulation Explorer \\
\textbf{GRACE} & Gravity Recovery And Climate Experiment   \\
\textbf{GRACE-FO} & Gravity Recovery And Climate Experiment -- Follow On  \\
\textbf{KBR} & K-Band Ranging \\
\textbf{LISA} & Laser Interferometer Space Antenna \\
\textbf{ll-SST} & Low-Low Satellite-to-Satellite Tracking   \\
\textbf{LPF} & Laser Interferometer Space Antenna -- Pathfinder \\
\textbf{LRI} & Laser Ranging Interferometer  \\
\textbf{MAGIC} & Mass change And Geosciences International Constellation  \\ 
\textbf{NASA} & National Aeronautics and Space Administration \\
\textbf{NGGM} & Next Generation Gravimetry Mission \\
\textbf{QACC} & Quantum Accelerometry \\
\textbf{SGRS} & Simplified Gravitational Reference Sensor \\
\textbf{SH} & Spherical Harmonic\\
\textbf{SuperSTAR} & Super Space Three-axis Accelerometer for Research \\
\textbf{TM}  & Test Mass\\
\textbf{TM-EH} & Test Mass -- Electrode Housing \\
\textbf{XHPS} & eXtended High Performance Satellite Dynamics Simulator
\end{tabular}
\end{table}

\end{document}